\documentclass[amsmath,amssymb,nofootinbib,notitlepage,superscriptaddress,twocolumn]{revtex4-2}
\pdfoutput=1

\usepackage[bottom]{footmisc}
\usepackage[colorlinks=true]{hyperref}
\usepackage[raggedright]{subfigure}
\usepackage{lipsum, babel}
\usepackage{comment}
\usepackage{slashbox}
\usepackage{amssymb}
\usepackage{tabularx}
\usepackage{cancel}
\usepackage{pifont}
\usepackage{soul}
\usepackage{mathtools,amsmath,amssymb,mathrsfs,color,esint}
\usepackage{array}
\usepackage{xcolor}
\newcolumntype{C}[1]{>{\centering\arraybackslash}p{#1}}
\newcommand{\p}{\partial}
\newcommand{\e}{\mathfrak{e}}

\begin{document}

\title{Accumulation of Charge on an Extremal Black Hole's Event Horizon}

\author{Zachary Gelles}
\email{zgelles@princeton.edu}
\affiliation{Department of Physics, Princeton University, Princeton, NJ 08540, USA}
\author{Frans Pretorius}
\email{fpretori@princeton.edu}
\affiliation{Department of Physics, Princeton University, Princeton, NJ 08540, USA}

\begin{abstract}
We numerically analyze the behavior of a charged scalar field on a fixed extremal Reissner-Nordstr\"om background. We find an extension of the Aretakis instability characterized by an accumulation of charge on the extremal event horizon. In particular, when the electromagnetic coupling to the scalar field is sufficiently large, the charge density on the horizon asymptotes to a nonzero constant at late times. By constructing monochromatic initial data at the onset of charged superradiance, we give evidence supporting the claim that this instability is connected to the presence of a nearly zero-damped mode. Throughout this work, we employ a numerical integration scheme in compactified double-null coordinates, which allows us to capture the asymptotic behavior of the matter at the boundaries of the spacetime. 
\end{abstract}

\maketitle

\section{Introduction\label{sec:intro}}
Black holes in nature are conjectured to be described by three numbers alone: mass $M$, electric charge $Q$, and angular momentum $J$. Extremal black holes --- which satisfy $J^2/M^2+Q^2=M^2$ --- have drawn considerable interest from physicists and mathematicians alike due to the myriad unique phenomena that arise in this limit. In particular, extremal black holes are not linearly stable to external perturbations, setting them apart from their more stable, sub-extremal counterparts. On a fixed background metric, radial derivatives of massless fields will not decay on the horizon of a maximally charged Reissner-Nordstr\"om (RN) black hole \cite{aretakis2011stability1,aretakis2011stability2} or a maximally rotating Kerr black hole \cite{aretakis_kerr} --- a result known as the Aretakis instability.

In order to properly understand the physical consequences of this horizon instability, one must analyze the fully nonlinear problem, evolving any external fields in conjunction with the underlying spacetime geometry. In general, and for the simplest scenario where the perturbation does not involve additional matter fields, this requires solving the coupled Einstein-Maxwell system of equations. This is a challenging problem, in particular since it cannot be addressed in spherical symmetry --- where the RN solution describes the relevant extremal black hole --- as neither the Einstein equations nor the Maxwell equations permit radiation then. As such, studying the stability of extremal RN black holes in the simplest possible setting requires one to couple to a matter field that has non-trivial dynamics in spherical symmetry.

In this paper, we will accordingly take a step back and aim to improve our understanding of the Aretakis instability in spherical symmetry. Specifically, we will numerically analyze an extremal RN black hole subject to perturbation by a spherically symmetric, electrically charged scalar field. 

Our work is intended to build towards an extension of that of Murata, Reall \& Tanahashi (MRT \cite{murata_what_2013}), who numerically investigated the non-linear evolution of an \emph{uncharged} scalar field in the presence of a RN black hole. Introducing an electromagnetic coupling to the scalar field may allow us to draw more explicit connections to the extremal-Kerr problem, in that a charged field will allow evolution of the charge $Q$ of the black hole, in analogy with gravitational radiation (or matter) allowing the angular momentum $J$ to evolve. In that sense, one may consider charged matter incident on an extremal RN black hole to serve as a toy model for rotating matter incident on an extremal Kerr black hole.

Charged scalar electrodynamics in the presence of a RN black hole is governed by three sets of equations: the wave equations for the components of the complex scalar field, Maxwell's equations for the electromagnetic field, and Einstein's equations for the spacetime geometry. The linearized problem for the scalar field --- fixing the background metric and electromagnetic potential --- was first investigated for general RN black holes by Hod \& Piran \cite{hod1998late_1,hod1998late_2,hod1998late_3} and specifically in the extremal limit by Zimmerman \cite{zimmerman_horizon_2017}. Zimmerman found that the charged scalar field experiences an enhanced Aretakis instability compared to that of the uncharged scalar field, with radial derivatives growing along the horizon instead of remaining constant. Zimmerman also demonstrated that this enhanced instability can indeed be mapped onto analogous results for linearized perturbations of Kerr \cite{casals_horizon_2016}. 

To date, however, the question of horizon (in)stability of a charged scalar field in spherical symmetry in the fully non-linear regime remains largely unexplored\footnote{Note that \cite{baake_superradiance_2016} did perform a fully non-linear investigation of scalar electrodynamics in curved spacetime, albeit without an analysis of horizon instabilities. Also, \cite{lucietti_horizon_2013} demonstrated that an Aretakis-like instability arises from electromagnetic perturbations of an extremal RN black hole, although it is not specific to the case of the charged scalar field.}. Our aim is to eventually tackle this problem. But since including back-reaction presents its own set of challenges in numerical analysis, as a first step here, we focus on evolving the coupled charged scalar/Maxwell system --- charged scalar electrodynamics (SED) --- on a fixed RN background geometry. Specifically, we implement a new finite difference code based on compactified, double-null coordinates to solve the corresponding partial differential equations. Such a coordinate system is advantageous in that it allows us to reach both the future horizon and future null infinity, ensuring that we capture the correct asymptotic behavior of the charged scalar field at the relevant boundaries of the spacetime.

In analyzing the numerical output, we find that charged SED exhibits a plethora of interesting phenomena that are not present in the linearized analysis of Zimmerman \cite{zimmerman_horizon_2017}. Indeed, we find that for sufficiently large electromagnetic coupling, the \emph{charge} density does not decay at late times on the extremal event horizon. This result bears striking resemblance to the uncharged Aretakis instability, for which the \emph{energy} density does not decay at late times on the extremal horizon \cite{lucietti_horizon_2013}. 
To help explain this similarity, we derive a novel gauge invariant formulation of Maxwell's equations from which the asymptotic behavior of the matter fields naturally follows. 

Finally, we explore the physical origins of the charged Aretakis instability by numerically evolving initial data consisting of a monochromatic wave with frequency precisely at the onset of superradiance. This allows us to test the arguments put forth by \cite{zimmerman_horizon_2017,casals_horizon_2016,richartz_synch_2} concerning the relationship between the enhanced Aretakis instability and a weakly damped mode that coincides with the onset of superradiance at extremality. We find that when an extremal black hole is driven with a wave at this frequency, the resulting scalar field amplitude on the horizon grows monotonically until the amplitude is large enough that the non-linearities of SED mitigate the growth.

The rest of this paper is organized as follows. In Section~\ref{sec:coordinates}, we provide an overview of the Reissner-Nordstr\"om spacetime and the double-null coordinates we use to cover it. In Section~\ref{sec:SED}, we explain how to write scalar electrodynamics as a well-posed initial value problem in spherical symmetry. Readers who are not interested in the details of the formalism can skip to Section~\ref{sec:results1}, where we present the results of our numerical simulations initialized with compactly supported data. In Section~\ref{sec:superradiance}, we explicitly analyze the connection between the Aretakis instability and the onset of superradiance. We give concluding remarks in Section~\ref{sec:conclusion}. In Appendices ~\ref{app:coords} and \ref{app:qlorenz}, we provide further details of the horizon-penetrating coordinate system and quasi-Lorenz gauge condition we use, respectively. In Appendix ~\ref{app:gaugeinveom}, we give a more complete derivation of our gauge-invariant formulation of SED in spherical symmetry.  The numerical scheme, together with convergence studies, is described in detail in Appendix~\ref{app:numerics}.

\section{Reissner-Nordstr\"om Black Holes\label{sec:coordinates}}
In this section we review relevant aspects of the RN geometry and describe the coordinate systems that we use to integrate and analyze the Maxwell-Klein-Gordon equations throughout the spacetime.

\subsection{Reissner-Nordstr\"om Metric}
\label{sec:RNsec}
The RN spacetime describes a spherically symmetric black hole of mass $M$ and electric charge $Q_0$. The line element in polar coordinates $(t,r,\theta,\phi)$ is\begin{align}
    ds^2&=-F(r)dt^2+\frac{1}{F(r)}dr^2+r^2d\Omega^2,
\end{align}
where\begin{align}\label{F_def}
    F(r)\equiv 1-\frac{2M}{r}+\frac{Q_0^2}{r^2}
\end{align}
and $d\Omega^2$ is the volume element on the two-sphere:\begin{align}
    d\Omega^2\equiv d\theta^2+\sin^2\theta d\phi^2.
\end{align}
This metric solves the Einstein equations sourced by the electromagnetic Coulomb potential:\begin{align}
\label{eq:staticA}
    A_{\rm RN}&=-\frac{Q_0}{r}dt.
\end{align}
The roots of $F(r)$ delineate the inner and outer event horizons of the RN black hole, both of which are null hypersurfaces\begin{align}
    F(r)=\frac{(r-r_+)(r-r_-)}{r^2},\quad r_\pm=M\pm\sqrt{M^2-Q_0^2}.
\end{align}
These two distinct horizons only exist when $|Q_0|<M$. When $|Q_0|>M$ there are no horizons, with the metric then describing a spacetime containing a naked singularity. Extremality corresponds to the case $Q_0=M$, wherein the inner and outer event horizons degenerate to a single marginally trapped surface. 
A convenient parameter to describe the deviation from extremality is the surface gravity\begin{align}
\label{eq:kappaeq}
    \kappa_+\equiv \frac{r_+-r_-}{2r_+^2},
\end{align}
which is zero when $Q_0=M$. 

\begin{figure}[h]
    \centering
    \includegraphics[width=0.5\textwidth]{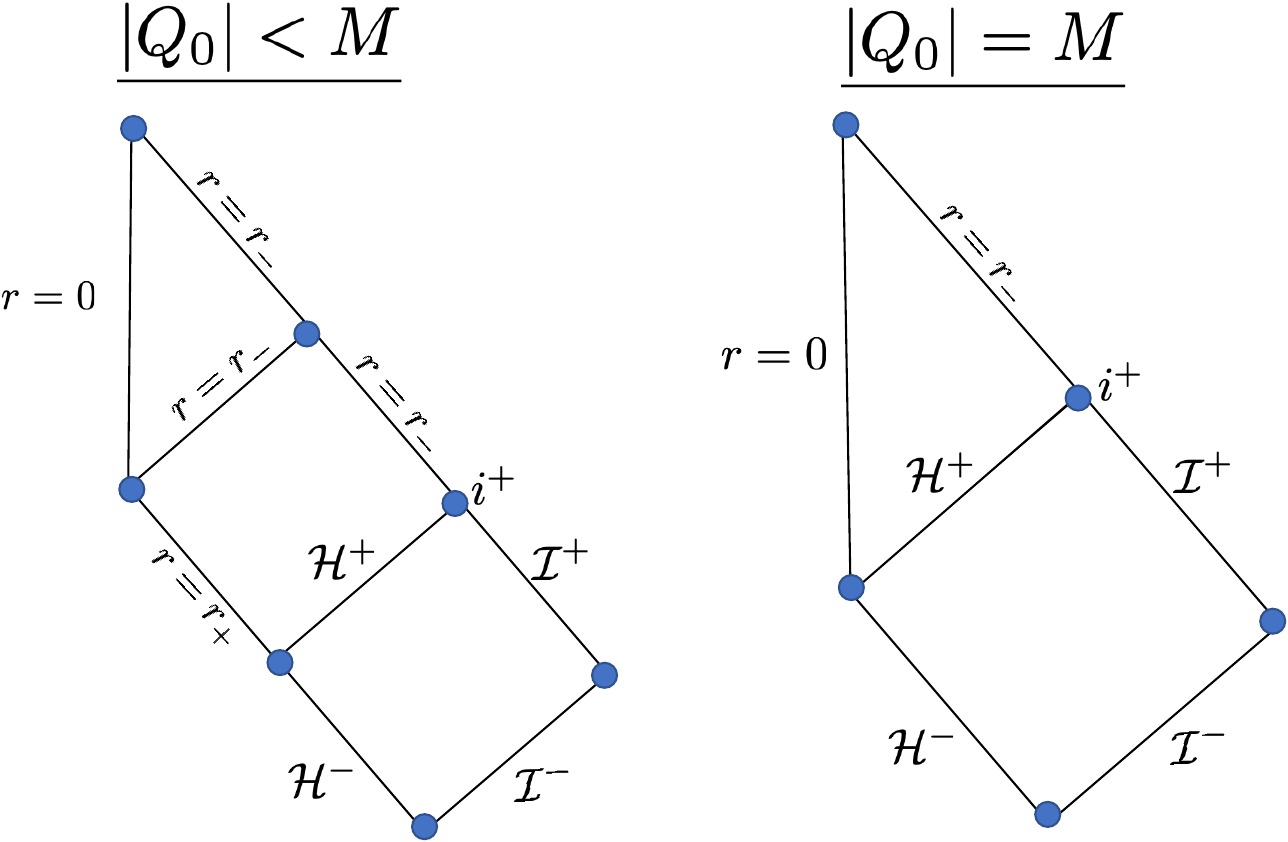}  
    \caption{Penrose diagram of RN black hole for the sub-extremal (left) and extremal (right) cases. The future/past horizons are denoted $\mathcal{H}^\pm$, future/past null infinities are denoted $\mathcal{I}^\pm$, and future timelike infinity is denoted by $i^+$. Both Penrose diagrams can be analytically continued to the future beyond the Cauchy horizon at $r=r_-$, and to the past beyond $\mathcal{H}^-$. Note that $r=0$ is a curvature singularity, and the blue circles denote coordinate singularities of the Penrose compactification (i.e., none of the black curves ``meet'' at these points in physical space).} 
    \label{fig:penrosediagram}
\end{figure}

Penrose diagrams for the sub-extremal and extremal RN black holes are shown in Figure~\ref{fig:penrosediagram}. In each diagram, the black hole exterior is bound by future/past null infinity $\mathcal{I}^\pm$ (which correspond to $r=\infty$) and the future/past horizon $\mathcal{H}^\pm$ (which correspond to $r=r_+$). 

\subsection{Double-Null Coordinates}
Throughout this work, we will describe the RN spacetime using null coordinates $U$ and $V$, in which the metric can be written as 
\begin{align}
\label{eq:uvmetric}
    ds^2&=-2f\,dU\,dV+r^2d\Omega^2
\end{align}
for some function $f(U,V)$. The metric function $r=r(U,V)$ remains the areal radius but now becomes a function of $U$ and $V$. A curve $U=\rm{const.}$ ($V={\rm const.}$) is an outgoing (ingoing) radial null geodesic, rendering these coordinates particularly well-adapted to describing the causal structure of the spacetime. 

 Null coordinates are not unique (i.e. rescaling $U$ by any function of itself, and likewise for $V$, preserves its null character), and different choices of $U$ and $V$ can come in handy for tackling different problems. In this work, we will rely on two particular choices of null coordinates. For analytic calculations, we will use Eddington-Finkelstein null coordinates, and for numerical integration we will use compactified Murata-Reall-Tanahashi coordinates. We describe each coordinate system below.

\subsubsection{Eddington-Finkelstein Null Coordinates}
The Eddington-Finkelstein (EF) null coordinates are defined by\begin{align}
    U_{\rm EF}&\equiv t-r_\star(r),\qquad V_{\rm EF}\equiv t+r_\star(r),
\end{align}
where the tortoise coordinate $r_\star$ satisfies\begin{align}
   \frac{dr_\star}{dr}=\frac{1}{F(r)},
\end{align} 
with $F(r)$ given by Eq.~\ref{F_def}.
The EF coordinates are a natural choice for a range of physical problems, in particular because they coincide with proper time of stationary observers far from the black hole ($r\rightarrow \infty$). The RN metric in these coordinates takes the form of Eq.~\ref{eq:uvmetric} with\begin{align}
    f\rightarrow f_{\rm EF}&=\frac{F(r)}{2},
\end{align}
again noting that $r=r(U_{\rm EF},V_{\rm EF})$.
The EF coordinates cover only the portion of the spacetime exterior to the black hole, with $\mathcal{H}^+$ ($\mathcal{H}^-$) at $U_{\rm EF}=\infty$ ($V_{\rm EF}=-\infty$), and $\mathcal{I}^+$ ($\mathcal{I}^-)$ at $V_{\rm EF}=\infty$ ($U_{\rm EF}=-\infty$).

While integrating to infinity is straightforward in analytic calculations, the same cannot be said for numerical evolution. In order to numerically investigate the behavior of charged scalar fields on the horizon and at null infinity, we need to solve the equations of motion in a coordinate system that reaches these hypersurfaces in finite null coordinate time. To this end, we use the compactified MRT coordinates below.

\subsubsection{Compactified MRT coordinates}
The Murata-Reall-Tanahashi (MRT) coordinates (introduced in \cite{murata_what_2013}) are implicitly defined by the relations\footnote{Our choice of coordinates differs from \cite{murata_what_2013} by a factor of 2. Also, \cite{murata_what_2013} did not use a horizon-penetrating $V$ coordinate.} \begin{align}
\label{eq:murataeq}
    U_{\rm EF}&=-2r_\star(r_+-U_{\rm MRT}/2),\quad V_{\rm EF}=2r_\star(r_++V_{\rm MRT}/2),
\end{align}
where $(r_+-U_{\rm MRT}/2)$ and $(r_++V_{\rm MRT}/2)$ are understood as the arguments of the $r_\star$ function.

These coordinates are identical to their EF counterparts at null infinity but cross the event horizon in finite time: the future horizon corresponds to $U_{\rm MRT}=0$, and the past horizon corresponds to $V_{\rm MRT}=0$. Thus, by implementing our numerical scheme in terms of MRT coordinates, we can integrate into the black hole interior, which will be crucial for analyzing the behavior of radial derivatives along the horizon, and hence the Aretakis instability. 

Indeed, $U_{\rm MRT}$ becomes tangent to the radial coordinate $r$ on the future horizon:\begin{align}
\label{eq:expo}
    \frac{\p}{\p r}\bigg|_{\mathcal{H}^+}=\frac{1}{2}e^{-\kappa_+V_{\rm EF}}\frac{\p}{\p U_{\rm MRT}}\bigg|_{\mathcal{H}^+},
\end{align}
where $\kappa_+$ is the surface gravity defined in Eq.~\ref{eq:kappaeq}.  The above equation reflects the exponential sensitivity of null geodesic trajectories near the sub-extremal horizon due to the redshift effect. The extremal case, however, has $\kappa_+=0$ and therefore does not experience this redshift.

To ensure that null infinity can be reached in finite time in our numerical code, we compactify the MRT coordinates using the inverse tangent function:\begin{align}
    u_{\rm MRT}\equiv \tan^{-1}(U_{\rm MRT}/M),\quad v_{\rm MRT}\equiv \tan^{-1}\left(V_{\rm MRT}/M\right),
\end{align}
respectively sending future and past null infinity to the finite coordinate values of $v_{\rm MRT}=+\pi/2$ and $u_{\rm MRT}=-\pi/2$.
Having access to null infinity in our numerical simulation is useful in that it allows us to measure the correct quasi-normal modes and late time power-law tails of the scalar field there (discussed in \S\ref{sec:powerlawtail}) without extrapolation. 
We note that a compactified hyperboloidal system was similarly introduced in \cite{Aretakis_2024_v2} to study (uncharged) scalar fields on extremal backgrounds.

In terms of the compactified MRT coordinates, the RN metric takes the form of Eq.~\ref{eq:uvmetric} with 
\begin{align}
\label{eq:muratametric}
 f\rightarrow f_{\rm MRT}&=\frac{F(r)\sec^2u_{\rm MRT}\sec^2v_{\rm MRT}}{2F(r_+-\tan u_{\rm MRT}/2)F(r_++\tan v_{\rm MRT}/2)}.
\end{align}
The above expression remains smooth across $r_+$ and is also continuous in the extremal limit  $Q_0\to M$ (see Appendix~\ref{app:coords} for details). 
While the term $\sec^2v_{\rm MRT}$ does blow up at future null infinity, we detail in Appendix~\ref{app:numerics} how we factor out this divergence to stably evolve all dynamical variables to $v_{\rm MRT}=\pi/2$. 

Throughout the rest of this paper, we will drop the ``MRT" subscripts and use the lower-case $\{u,v\}$ to refer to compactified MRT coordinates. Unless explicitly stated otherwise, upper-case $\{U,V\}$ will be used to refer to arbitrary double-null coordinates.

\section{Scalar Electrodynamics \label{sec:SED}}
In this section, we outline the theory of scalar electrodynamics, starting first with a generic overview before describing how we can write the theory as a well-posed initial value problem.
\subsection{Overview}
\subsubsection{Covariant Description}
Scalar electrodynamics (SED) is a field theory used to model the interactions of electrically charged matter. In SED, the matter field is taken to be a complex scalar $\phi$, which couples to the electromagnetic gauge field $A_\mu$ via a coupling constant $\e$. In curved spacetime with a fixed background metric, the action is given by\footnote{We work in Gaussian units, where the factor of $\frac{1}{4\pi}$ is included in the Lagrangian (e.g. \cite{hod_critical_1997,MTW_book,Wald_book}). In these units, Coulomb's law is $E=Q/r^2$, and extremality of the RN solution occurs at $Q_0=M$. Additionally, note that we dropped the symmetry factor of 1/2 from the Lagrangian of \cite{hod_critical_1997}, as a complex field often does not have this symmetry factor under canonical normalization (see, e.g. \cite{PS_1995}).}\begin{align}
\label{eq:action}
    S&=\int\sqrt{-g}\,d^4x\left[-\frac{1}{16\pi}\mathcal{F}_{\mu\nu}\mathcal{F}^{\mu\nu}-(D_\mu\phi)(D^\mu\phi)^\star\right],
\end{align}
where $\sqrt{-g}$ is the metric determinant, the Faraday tensor is\begin{align}
    \mathcal{F}_{\mu\nu}&=\nabla_\mu A_\nu-\nabla_\nu A_\mu=\p_\mu A_\nu-\p_\nu A_\mu,
\end{align}
and the gauged covariant derivative is\footnote{The charge coupling in the covariant derivative has a minus sign, as appropriate for mostly positive signature and agreeing with the equations of motion in, e.g., \cite{konoplya_massive_2002,richartz_quasinormal_2014}.}\begin{align}
      D_\mu\equiv\nabla_\mu-i\e A_\mu.
\end{align}
This theory has a $U(1)$ gauge symmetry that leaves the action invariant under the transformation\begin{align}
    \phi\to  e^{i\e\alpha(x)}\phi,\qquad A_\mu\to A_\mu+\p_\mu\alpha(x).
\end{align}
As we will see, the existence of this gauge symmetry allows us to analyze our results in a manner distinct from that of the uncharged field theory, providing clearer insights into many aspects of the problem. 

Variation of the action with respect to the fields $\{A_{\mu},\phi,\phi^\star\}$ gives the equations of motion:\begin{align}
    &\nabla_{\nu}\mathcal{F}^{\mu\nu}=4\pi J^\mu\,\,\text{(Maxwell)}\label{eq:maxwell1}\\
    &D_\mu D^\mu\phi=0\,\,\text{(Wave)}\label{eq:wave1}\\
    &(D_\mu D^\mu\phi)^\star=0\,\,\text{(Wave Conjugate)}\label{eq:wave2},
\end{align}
where the electromagnetic current is\begin{align}
\label{eq:currentdef}
    J^\mu=i\e[\phi (D^\mu\phi)^\star-\phi^\star (D^\mu\phi)]=-2\e\,{\rm Im}[\phi (D^\mu\phi)^\star].
\end{align}
From these fields, one can build the stress tensor for SED\begin{align}
    T^{\mu\nu}&=\frac{1}{4\pi}\mathcal{F}_{\mu\rho}\mathcal{F}^{\,\,\rho}_\nu-\frac{1}{16\pi}g_{\mu\nu}\mathcal{F}^{\alpha\beta}\mathcal{F}_{\alpha\beta}\\&\nonumber+(D_\mu\phi)(D_\nu\phi)^\star+(D_\nu\phi)(D_\mu\phi)^\star-g_{\mu\nu}(D_\alpha\phi)(D^\alpha\phi)^\star,
\end{align}
which is conserved in the sense that $\nabla_\mu T^{\mu\nu}=0$ when the equations of motion are satisfied.

Part of this stress tensor is sourced by the point charge at $r=0$, which we treat as static in this work. It will therefore be helpful for us to isolate the purely dynamical portion of the stress tensor, which is given by \begin{align}
    \mathcal{T}^{\mu\nu}\equiv T^{\mu\nu}-T^{\mu\nu}_{\rm RN},
\end{align}
where $T^{\mu\nu}_{\rm RN}$ is the RN stress tensor sourced only by the vector potential of Eq.~\ref{eq:staticA}. Since $T^{\mu\nu}$ and $T^{\mu\nu}_{\rm RN}$ are both conserved by default, then so too is $\mathcal{T}^{\mu\nu}$.

From here, let us now break down the components of $\mathcal{F}$, $J$, and $\mathcal{T}$ in terms of double-null coordinates.

\subsubsection{Fields in Double Null Coordinates}
In spherical symmetry there is no magnetic field, and only the radial component of the electric field is nonzero. For any two null coordinates $U$ and $V$, this single component of the electric field can be encoded into the Faraday tensor as\begin{align}
\label{eq:faradaydef}
    \mathcal{F}&=-\frac{Qf}{r^2}dU\wedge dV,
\end{align}
where $Q=Q(U,V)$ represents the total charge interior to $r(U,V)$, and $f$ is defined in Eq.~\ref{eq:uvmetric}. As one can readily check, the above equation satisfies\begin{align}\label{eq:coul}
   \mathcal{F}^{\mu\nu}\mathcal{F}_{\mu\nu}=-\frac{2Q^2}{r^4},
\end{align}
which in this spherically symmetric situation effectively reduces to Coulomb's law\footnote{To see why the minus sign is included in Eq.~\ref{eq:faradaydef}, consider Minkowski space with $U=t-r$, $V=t+r$, and $f=1/2$. Then, Eq.~\ref{eq:faradaydef} will correctly reduce to $F^{tr}=+Q/r^2$.}.

From here, one can re-cast Maxwell's equations in terms of $Q$. Expanding Eq.~\ref{eq:maxwell1}, we obtain\begin{align}
\label{eq:maxwellcons1}
    Q_{,U}&=4\pi r^2J_U
    \\
    \label{eq:maxwellcons2}
    Q_{,V}&=-4\pi r^2J_V,
\end{align}
where commas denote partial differentiation. The current components $J_U$ and $J_V$ are most easily expressed in terms of the renormalized fields\begin{align}
    \overline{\xi}\equiv {\rm Re}(r\phi),\qquad \overline{\Pi}\equiv {\rm Im}(r\phi),
\end{align}
which will remain nonzero at null infinity since the scalar field decays as $r^{-1}$. This gives\begin{align}
    r^2J_\mu=-2\e[\overline{\xi}_{,\mu}\overline{\Pi}-\overline{\xi}\overline{\Pi}_{,\mu}+\e A_\mu(\overline{\xi}^2+\overline{\Pi}^2)],\label{eq:currentdef2}
\end{align}
which can be plugged directly into the right hand side of Eqs.~\ref{eq:maxwellcons1}-\ref{eq:maxwellcons2}.

The form of Maxwell's equations derived above serves as a constraint on the evolution of $Q$:  Eqs.~\ref{eq:maxwellcons1}-\ref{eq:maxwellcons2} must be obeyed along each null hypersurface of the spacetime. We stress that while $Q$ is dynamical, we treat $Q_0$ --- the charge of the background RN spacetime --- as a constant in this work.

Finally, we can expand the components of the dynamical stress tensor in terms of the gauge-invariant quantities $Q$ and $P\equiv |r\phi|$ as\begin{align}
\label{eq:tdef}
    \mathcal{T}_{UU}&=\frac{2}{r^2}\left[\left(\frac{Q_{,U}}{8\pi \e P}\right)^2+r^2\left(\left(\frac{P}{r}\right)_{,U}\right)^2\right]
    \\
   \nonumber  \mathcal{T}_{VV}&=\frac{2}{r^2}\left[\left(\frac{Q_{,V}}{8\pi \e P}\right)^2+r^2\left(\left(\frac{P}{r}\right)_{,V}\right)^2\right]
     \\
   \nonumber  \mathcal{T}_{UV}&=\frac{f(Q^2-Q_0^2)}{8\pi r^4},
\end{align}
which are all well-defined so long as $\e \neq 0$.

\subsection{Initial Value Problem}
Let us now take the above decomposition and use it to construct a well-posed initial value problem for SED in double-null coordinates. To do so, we will need to begin by fixing a gauge for $A_\mu$. We can then use this gauge choice to construct initial data that can then be stably evolved in $U$ and $V$.
\subsubsection{Gauge Fixing}
Fixing a gauge for $A_\mu$ is essential to formulate a well-posed system of evolution equations. A common choice in the literature is Lorenz gauge $\nabla_\mu A^\mu=0$, for which a static point charge is described by Eq.~\ref{eq:staticA}. 

However, Lorenz gauge will \emph{not} be ideal for our numerical integration scheme. Indeed, the vector potential in Lorenz gauge decays as $r^{-1}$, which means its integral will not be finite at null infinity. To ensure that $A_\mu$ decays more sharply, we introduce ``Quasi-Lorenz Gauge," which we defined as\begin{align}
\label{eq:quasilorenz}
    \nabla_\mu(r^{-2}A^\mu)=0.
\end{align}
This gauge condition is particularly well-adapted to spherical symmetry, where in double-null coordinates it takes the simple form
\begin{align}
\label{eq:quasiL}
    A_{U,V}+A_{V,U}=0.
\end{align}
Combining this gauge choice with compactified MRT coordinates, the vector potential for a RN black hole of charge $Q_0$ is given by (see Appendix~\ref{app:qlorenz} for a derivation):\begin{align}
\label{eq:gaugequasi1}
    A_u(u,v)&=
        \frac{Q_0\sec^2u}{2F(r_+-\tan u/2)}\left[\frac{1}{r(u,v_0)}-\frac{1}{r(u,v)}\right]
    \\
    \label{eq:gaugequasi2}
    A_v(u,v)&=\frac{Q_0\sec^2v}{2F(r_++\tan v/2)}\left[\frac{1}{r(u_0,v)}-\frac{1}{r(u,v)}\right],
\end{align}
where $(u_0,v_0)$ is an arbitrary point in the spacetime that defines the boundary condition $A_\mu(u_0,v_0)=0$.

In Appendix~\ref{app:qlorenz}, we show that these expressions for $A_u$ and $A_v$ remain smooth across the event horizon and remain finite at null infinity, in contrast to their Lorenz-gauge counterparts. Therefore, Quasi-Lorenz gauge is best suited for our numerical integration scheme.

\subsubsection{Initial Data\label{sec:initdata}}
The construction of initial data in double-null coordinates is different than the ``standard" procedure one would adopt when integrating in $t$ and $r$. When integrating in $U$ and $V$,  one begins by choosing an origin point $(U_0,V_0)$ in the spacetime. Initial data for all dynamical fields is then pasted along the future null cone of this origin, which corresponds to the ingoing and outgoing null hypersurfaces $\mathcal{N}_{A}=\{V=V_0,U\geq U_0\}$ and $\mathcal{N}_{B}=\{U=U_0,V\geq V_0\}$ \cite{burko_late-time_1997,murata_what_2013}. Throughout this work, define the origin in EF coordinates as \begin{align} 
    (U_{0,\rm EF},V_{0,\rm EF})&=(0,0),
\end{align}
which can be mapped to MRT coordinates for a given $Q_0$ via Eq.~\ref{eq:murataeq}.
Additionally, we will consider only non-trivial ingoing initial data (i.e. a scalar field profile supported only on $\mathcal{N}_{B}$).

To ensure that the initial data obeys Maxwell's constraint equations (Eqs.~\ref{eq:maxwellcons1}-\ref{eq:maxwellcons2}) and the Quasi-Lorenz gauge condition (Eq.~\ref{eq:quasiL}), one can adopt the following four-step procedure, which we implement explicitly in our numerical code:
\begin{enumerate}
    \item Initialize the scalar field. In this work, we will take the scalar field to consist of an ingoing, monochromatic\footnote{For the purposes of comparing to past work, we assume that $\tilde{\omega}$ is the frequency in Lorenz gauge. For our numerical simulations, we then subsequently rotate $\phi$ by the appropriate local phase to convert to quasi-Lorenz gauge.} wave of frequency $\tilde{\omega}$ tapered by a smooth envelope $\mathcal{A}(V)$:\begin{align}
        (\overline{\xi}+i\overline{\Pi})|_{\mathcal{N}_{A}}&=0
        \\
    (\overline{\xi}+i\overline{\Pi})|_{\mathcal{N}_{B}}&=\mathcal{A}(V)e^{-i\tilde{\omega} V}\label{eq:ampmod},
    \end{align}
    where $\mathcal{A}(V)$ is real and non-negative.

    \item Set $A_U|_{\mathcal{N}_{A}}=A_V|_{\mathcal{N}_{B}}=0$, which is just a boundary condition on the vector potential.

    \item Integrate Maxwell's constraint equations (Eqs.~\ref{eq:maxwellcons1}-\ref{eq:maxwellcons2}) to determine the total charge $Q$ on each hypersurface. For the ingoing perturbation outlined above, Eq.~\ref{eq:maxwellcons1} trivially integrates to\begin{align}
        Q|_{\mathcal{N}_{A}}&=Q_0,
    \end{align} 
    while Eq.~\ref{eq:maxwellcons2} integrates to:\begin{align}
    \label{eq:initcharge}
    Q|_{\mathcal{N}_{B}}&=Q_0+8\pi \e\tilde{\omega} \int_{V_0}^{V}dV'\,\mathcal{A}(V')^2.
    \end{align}
    
    \item Impose Quasi-Lorenz gauge to obtain the remaining component of the gauge field on each hypersurface. For the ingoing perturbation, $A_{V}|_{\mathcal{N}_{A}}$ is unaffected by the matter and will therefore be given by Eq.~\ref{eq:gaugequasi2}, whereas $A_{U}|_{\mathcal{N}_{B}}$ must be computed by definite integration:
    \begin{align}
    A_{U}|_{\mathcal{N}_{B}}&=\int_{V_0}^VdV'\frac{fQ}{2r^2}.
    \end{align}
    
\end{enumerate}

The $\tilde{\omega}$ parameterization (Eq.~\ref{eq:ampmod}) is particularly useful because $\tilde{\omega}$ has a clear physical interpretation: positive (negative) $\tilde{\omega}$ produces positive (negative) charge density on $\mathcal{N}_{B}$.
When $\mathcal{A}$ is a constant and the data is purely monochromatic, we can relate $\tilde{\omega}$ to a charge-to-mass ratio for the initial data. To see this, consider the following measure of the mass accreting onto the system from the matter fields on $\mathcal{N}_{B}$:\begin{align}
\label{eq:Macc}
    M_{\rm acc}|_{\mathcal{N}_{B}}&\equiv\int_{\mathcal{N}_{B}}\mathcal{T}^\mu_{\;\;t}d\Sigma_\mu,
\end{align}
which corresponds to the Killing energy density a stationary observer would measure. Here, $d\Sigma_\mu$ is the vector normal to $\mathcal{N}_B$ that is appropriately normalized for a null hypersurface (see Chapter 3 of \cite{poisson_toolkit} for an in depth discussion of this matter). For data supported far from the black hole, one can plug this into the stress tensor from Eq.~\ref{eq:tdef} and expand in powers of $r^{-1}$ to find\begin{align}
    \frac{d M_{\rm acc}}{dV_{\rm EF}}\bigg|_{\mathcal{N}_{B}}&=4\pi r^2\mathcal{T}_{V_{\rm EF}V_{\rm EF}}+\mathcal{O}(r^{-1})\\\label{eq:dmdveq}&=8\pi\left[\left(\frac{Q_{,V_{\rm EF}}}{8\pi \e P}\right)^2+(P_{,V_{\rm EF}})^2\right]+\mathcal{O}(r^{-1}).
\end{align}

Assuming a constant amplitude $\mathcal{A}$, then $P_{,V_{\rm EF}}=0$ and so we can plug in Eq.~\ref{eq:initcharge} to find\begin{align}
    \nonumber\frac{dM_{\rm acc}}{dV_{\rm EF}}\bigg|_{\mathcal{N}_{B}}&=8\pi\tilde{\omega}^2\mathcal{A}^2+\mathcal{O}(r^{-1})
    \\
    \label{eq:chargetomass}
    \Longrightarrow\frac{dQ}{dM_{\rm acc}}\bigg|_{\mathcal{N}_{B}}&=\frac{\e}{\tilde{\omega}}+\mathcal{O}(r^{-1}).
\end{align}
Thus, the quantity $\e/\tilde{\omega}$ is directly interpreted as a charge-to-mass ratio for the initial data in this large $r$, monochromatic regime.

Eqs.~\ref{eq:initcharge} and \ref{eq:dmdveq} are also useful for estimating the conditions under which non-linearities are expected to remain small at early times. In particular, metric backreaction can be reliably ignored at early times when $M_{\rm acc}\ll 1$, and charge backreaction (i.e. electric self-force) can be reliably ignored at early times when $|Q-Q_0|\ll Q_0$. If $\mathcal{A}(V)$ can be approximated as a Gaussian of amplitude $\mathcal{A}_0$ and width $\sigma$, then these two conditions respectively translate to\begin{align}
   \label{eq:mestimate} 4\pi^{3/2}\mathcal{A}_0^2(\sigma^{-1}+2\tilde{\omega}^2\sigma)&\ll 1\quad&& (\text{Einstein})
    \\
    8\pi^{3/2}\e\tilde{\omega}\mathcal{A}_0^2\sigma&\ll Q_0\quad&&(\text{Maxwell})\label{eq:qestimate}.
\end{align}
We always expect metric backreaction to be relevant at late times, especially near extremality where instabilities become dynamically important \cite{murata_what_2013}. However, Eqs.~\ref{eq:mestimate} and \ref{eq:qestimate} are useful heuristics for determining the time-scales in which we expect linear approximations to remain valid.

\subsubsection{Evolution\label{sec:evolution}}
\label{sec:eom}
With the initial data constructed along $\mathcal{N}_{A}$ and $\mathcal{N}_{B}$, we can evolve the fields everywhere to the causal future of the point $(U_0,V_0)$. To do so, let us write down a system of well-posed evolution equations in double-null coordinates. First, the wave equations (Eqs~\ref{eq:wave1}-\ref{eq:wave2}) in Quasi-Lorenz gauge become\begin{align}
\label{eq:waveeqn}
    &\overline{\xi}_{,UV}=\overline{\xi}g+\e^2\overline{\xi}A_UA_V-\e\left(A_V\overline{\Pi}_{,U}+A_U\overline{\Pi}_{,V}\right)
    \\
    &\overline{\Pi}_{,UV}=\overline{\Pi}g+\e^2\overline{\Pi}A_UA_V
    +\e(A_V\overline{\xi}_{,U}+A_U\overline{\xi}_{,V}),
\end{align}
where\begin{align}
\label{eq:gdefn}
     g(U,V)&\equiv \frac{r_{,UV}}{r}=-\frac{fF'(r)}{2r}.
\end{align}
 The equations for both $\overline{\xi}$ and $\overline{\Pi}$ are manifestly hyperbolic.  
 
By differentiating Eqs.~\ref{eq:maxwellcons1}-\ref{eq:maxwellcons2} and applying the Quasi-Lorenz gauge condition, one can also derive hyperbolic evolution equations for the gauge field:\begin{align}
    A_{U,UV}&=\frac{f}{r^2}\left[2\pi r^2J_U-\left(\frac{r^2}{f}\right)_{,U}A_{U,V}\right]
    \\
    \label{eq:Aeq}
    A_{V,UV}&=\frac{f}{r^2}\left[2\pi r^2J_V-\left(\frac{r^2}{f}\right)_{,V}A_{V,U}\right].
\end{align}
Eqs.~\ref{eq:waveeqn}-\ref{eq:Aeq} thus describe the evolution of the four fields $\{\overline{\xi},\overline{\Pi},A_U,A_V\}$ throughout the spacetime. These equations of motion are valid for any choice of double-null coordinates, so long as one plugs in the appropriate expression for $f$.

We recall, however, that these four fields are \emph{not} independent due to the redundancy introduced by the $U(1)$ gauge symmetry. Indeed, scalar electrodynamics in spherical symmetry possesses only two physically meaningful degrees of freedom\footnote{And technically only one {\em radiative} degree of freedom due to the Maxwell constraint equations.} : the scalar amplitude $P\equiv |r\phi|$, and the enclosed charge $Q$ (equivalently, the radial electric field). In Appendix~\ref{app:gaugeinveom}, we show that the four equations of motion in Eqs.~\ref{eq:waveeqn}-\ref{eq:Aeq} can accordingly be condensed into two. They are\begin{align}
    &\label{eq:peq}P_{,UV}=Pg-\frac{Q_{,U}Q_{,V}}{64\pi^2\e^2P^3}
    \\
 &\label{eq:qeq}Q_{,UV}=\frac{P_{,U}Q_{,V}}{P}+\frac{P_{,V}Q_{,U}}{P}-\frac{4\pi \e^2fP^2Q}{r^2} .
\end{align}
So, after posing self-consistent initial data using Eqs ~\ref{eq:maxwellcons1}-\ref{eq:maxwellcons2}, one can technically solve for the future evolution of SED in spherical symmetry using only the gauge-invariant quantities $P$ and $Q$.

However, these gauge-invariant evolution equations are more challenging to solve numerically due to the presence of powers of $P$ in the denominator of terms in Eqs.~\ref{eq:peq},~\ref{eq:qeq} (i.e., in electro-vacuum where no scalar field is present, the corresponding $\frac{0}{0}$ terms would require careful treatment for stable evolution). Therefore, in practice we use Eqs.~\ref{eq:waveeqn}-\ref{eq:Aeq} for numerical evolution and then separately compute the gauge-invariant quantities\begin{align}
    P&=\sqrt{\overline{\xi}^2+\overline{\Pi}^2},\quad Q=\frac{r^2}{f}(A_{U,V}-A_{V,U}).
\end{align}
Details of the numerical evolution scheme and code implementation are described at length in Appendix~\ref{app:numerics}.

In the next two sections, we describe our results numerically evolving several sets of initial data. First, we explore the evolution of compactly supported initial data, which sheds light on the nature of the Aretakis instability in the presence of a dynamical electromagnetic field. Second, we explore the evolution of purely monochromatic initial data (and hence not compactly supported), which sheds light on the connection between the Aretakis instability and a single weakly damped mode connected to superradiance.

\section{Results - Compactly Supported Initial Data \label{sec:results1}} 
In this section, we use our numerical code to analyze scalar electrodynamics with ingoing, compactly supported initial data. After evolving the initial data, we will analyze the asymptotic dynamics of the scalar field, as well as the role of the dynamical Maxwell field in enhancing the Aretakis instability.

To construct compactly supported initial data, we employ the procedure outlined in \S\ref{sec:initdata}, taking the scalar envelope $\mathcal{A}$ (defined in Eq.~\ref{eq:ampmod}) to be a bump function of width $15M$ and amplitude 0.01 (analogous to the initial data used in \cite{murata_what_2013}):
\begin{widetext}
\begin{align}
\label{eq:initdatapulse}
    \mathcal{A}(V)&=\begin{cases}0.01\exp\left[\frac{15M}{4}
    \left(\frac{1}{V_{\rm EF}-15M}-\frac{1}{V_{\rm EF}}\right)+1\right],&0\leq V_{\rm EF}\leq 15M
    \\
    0,&{\rm else}.
    \end{cases}
\end{align}
\end{widetext}
As one can check, the function $\mathcal{A}(V)$ is smooth everywhere and is supported only for $0\leq V_{\rm EF}\leq 15M$. Furthermore, this choice of amplitude and width ensures that the conditions in Eqs.~\ref{eq:mestimate} and \ref{eq:qestimate} are met.

We then finish the construction of the initial data and run the simulation for two values of the background charge $Q_0$ (sub-extremal, extremal); three values of the charge coupling $\e$ (uncharged, weakly charged, strongly charged); and four values of the frequency $\tilde{\omega}$ (zero, small, medium, large):\begin{align}
\label{eq:qpar}
   Q_0/M&\in\{0.999,1.0\}
   \\
   \label{eq:epar}
   \e Q_0 &\in \{0,0.4,0.6\}
   \\
   \label{eq:omegapar}
   \tilde{\omega}&\in\{0,\e/2,\e,3\e/2\}.
\end{align}
In the next subsection, we will explain why these values of $\e Q_0$ and $\tilde{\omega}$ qualify as relatively weak/small and strong/large. Throughout the rest of this work, we will also set $M=1$.

\begin{figure*}[t]
    \centering
    \includegraphics[width=0.99\textwidth]{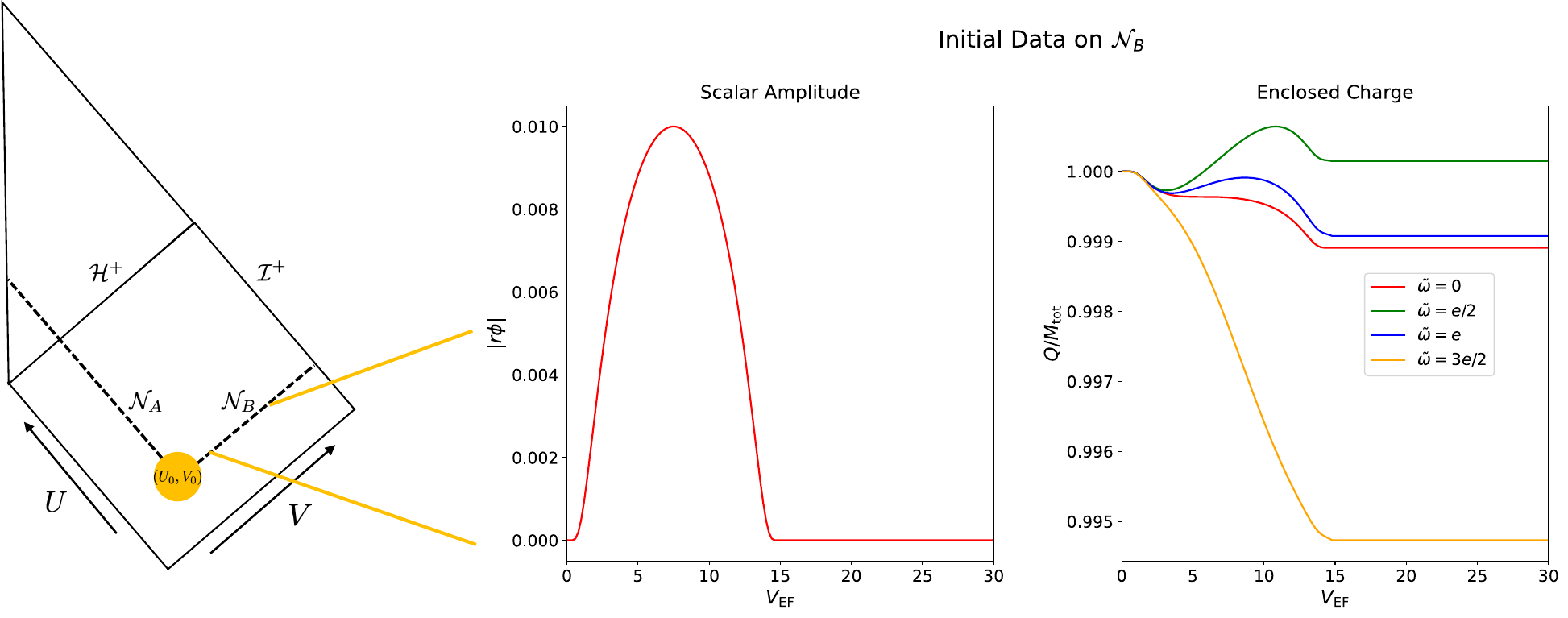}  
    \caption{Ingoing initial data for $Q_0=1.0,\,\e Q_0=0.6$. The scalar amplitude is the same for all choices of parameters and is shown in the middle panel. In the right panel, sample charge-to-mass ratios $Q/M_{\rm tot}$ are shown for three different values of $\tilde{\omega}$. The denominator of this ratio is $M_{\rm tot}=1+M_{\rm acc}$, where $M_{\rm acc}$ is defined in Eq.~\ref{eq:Macc}.} 
    \label{fig:initdatafig}
\end{figure*}

Since $\mathcal{A}$ is not constant along this compactly supported pulse, $\tilde{\omega}$ loses its interpretation as charge-to-mass ratio as given by Eq.~\ref{eq:chargetomass}. However, as illustrated in the right panel of Figure~\ref{fig:initdatafig}, we still find that the initial data case with $\e/\tilde{\omega}<1$ has decreasing $Q/M_{\rm tot}$ along $\mathcal{N}_{B}$, while the initial data case with $\e/\tilde{\omega}>1$ has increasing $Q/M_{\rm tot}$ along $\mathcal{N}_{B}$; the exact case of $\tilde{\omega}=0$ gives constant $Q=Q_0$ on the initial data surface (see Eq.~\ref{eq:initcharge}) and thus produces monotonically decreasing $Q/M_{\rm tot}$. The middle panel of Figure~\ref{fig:initdatafig} plots the initial data for the scalar amplitude $|r\phi|$, which does not depend on any of the simulation parameters listed above.

For each choice of parameters, we perform the numerical integration in compactified MRT coordinates on a non-uniform mesh. Specifically, we integrate along hypersurfaces of constant $U$ from $u_{0}$ to $u_{f}=+1.0$, thus capturing the entire exterior as well as a significant portion of the interior. Our canonical simulation resolution features $10000\,(U)\times 20000\,(V)$ cells in the black hole exterior, with details of the numerical scheme and convergence studies presented in Appendix~\ref{app:numerics}.

As a first step in analyzing the numerical output, we begin by examining the evolution of the scalar field on the future horizon and at null infinity, which together bound the relevant portion of the black hole exterior. The results are presented in the following subsection, and the late-time scalings of the dynamical fields are summarized in Table~\ref{table:table1}.

\begin{table*}[t!]
\centering
\begin{tabularx}{\textwidth}{|X||X|X|X|X|X|X|} 
 \hline
 \multicolumn{1}{|X||}{} & 
 \multicolumn{1}{>{\centering\arraybackslash}X|}{$|Q|<1,\e=0$} & 
 \multicolumn{1}{>{\centering\arraybackslash}X|}{$|Q|<1, \e\neq 0$} & 
 \multicolumn{1}{>{\centering\arraybackslash}X|}{$|Q|=1,\e=0$} & 
 \multicolumn{1}{>{\centering\arraybackslash}X|}{$|Q|=1,\e=0,$ $H_0\neq 0$} & 
 \multicolumn{1}{>{\centering\arraybackslash}X|}{$|Q|=1,$ $ \e\neq 0$} \\ 
 \hline
  $|r\phi|$ on the Future Horizon & $V_{\rm EF}^{-3}$\qquad\qquad\qquad (Price \cite{price_nonspherical_1972}) & $V_{\rm EF}^{-2s}$\qquad\qquad\qquad (Hod \& Piran \cite{hod1998late_2}) & $V_{\rm EF}^{-2}$\qquad\qquad\qquad (Lucietti+ \cite{lucietti_horizon_2013}; Angelopoulos+ \cite{angelopoulos2020late}) & $V_{\rm EF}^{-1}$ \qquad\qquad\qquad (Lucietti+ \cite{lucietti_horizon_2013}; Angelopoulos+ \cite{angelopoulos2020late}) & $V^{-s}$ \qquad\qquad\qquad (Zimmerman \cite{zimmerman_horizon_2017}) \\ 
 \hline
 $|r\phi|$ at Null Infinity & $U_{\rm EF}^{-2}$\qquad\qquad\qquad (Gundlach+ \cite{gundlach1994late}) & $U_{\rm EF}^{-s}$ \qquad\qquad\qquad (Hod \& Piran \cite{hod1998late_2}) & $U_{\rm EF}^{-2}$\qquad\qquad\qquad (Angelopoulos+ \cite{angelopoulos_horizon_2018}) & $U_{\rm EF}^{-2}$ \qquad\qquad\qquad (Angelopoulos+ \cite{angelopoulos2020late}) & \textcolor{blue}{$U_{\rm EF}^{-s}$} \\ 
 \hline
  Energy Density on the Future Horizon & $V_{\rm EF}^{-6}$ \qquad\qquad\qquad (Follows from Price \cite{price_nonspherical_1972})& $V_{\rm EF}^{-4s}$\qquad\qquad\qquad(follows from Hod \& Piran \cite{hod1998late_2}) & $V_{\rm EF}^{-4}$ \qquad \qquad  (Lucietti+ \cite{lucietti_horizon_2013}) & $V_{\rm EF}^0$ (Lucietti+ \cite{lucietti_horizon_2013}) & \textcolor{blue}{$V_{\rm EF}^{2-2s}$} \\
  \hline 
  Charge Density on the Future Horizon & N/A & \textcolor{blue}{$V_{\rm EF}^{-4s}$} & N/A & N/A & \textcolor{blue}{$V_{\rm EF}^{1-2s}$} 
  \\
  \hline
\end{tabularx}
\caption{Decay of physical quantities on horizon and null infinity, with references for prior results indicated in parentheses. Our novel results are shown in blue. These scalings apply to compactly supported initial data (the uncharged ($\e=0$) extremal limit with non-compact initial data is summarized in \cite{aretakis_dynamics_2018}). The quantity $s\equiv {\rm Re} \left(\frac{1}{2}+\sqrt{\frac{1}{4}-(\e Q_0)^2}\right)\in \left[\frac{1}{2},1\right]$ was first introduced in \cite{hod1998late_2} and is equal to $\frac{1}{2}$ for all $|\e Q_0|>\frac{1}{2}$. $H_0$ is an initial transverse gradient of the scalar field across the horizon (Eq.~\ref{H0_def}), which is preserved during evolution when $|Q_0|=1$ and $\e=0$ \cite{aretakis2011stability1,aretakis2011stability2}.}
\label{table:table1}
\end{table*}

\subsection{Asymptotic Behavior of the Scalar Field}
\begin{figure*}[t]
    \centering
    \includegraphics[width=0.99\textwidth]{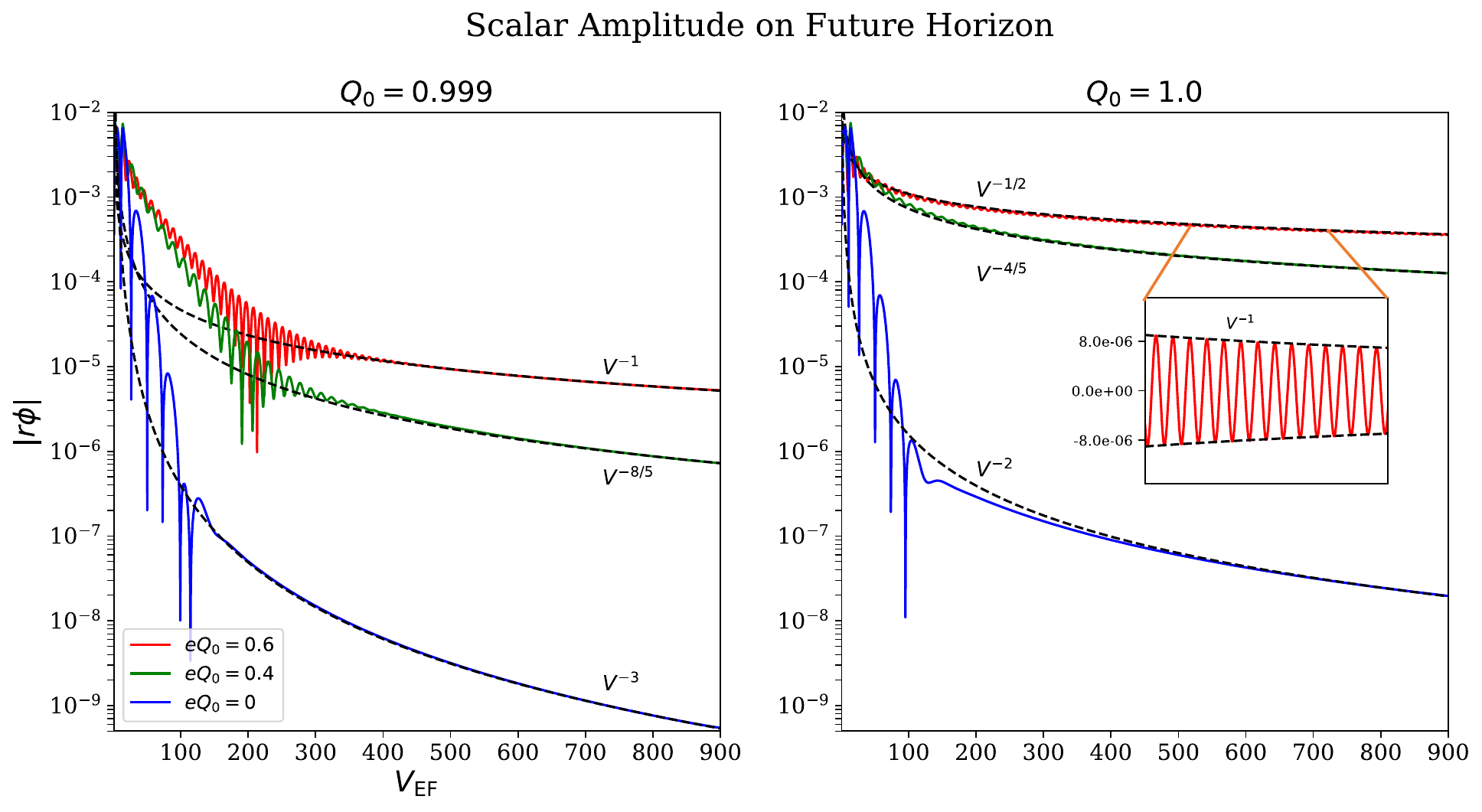}  
    \caption{Amplitude of the scalar field $|r\phi|$ on the future horizon $(u=0)$. Initial data is constructed to be neutral $(\tilde{\omega}=0)$, though evolution generically causes charge separation when $\e Q_0\neq 0$. The left panel shows results for the sub-extremal background, with the right panel showing results for the extremal background. Dashed lines indicate the approximate scalings for each power-law tail, fit by-eye, and guided by analytical expectations where known (see Table \ref{table:table1}, where $\e Q_0=(0,0.4,0.6)$ gives $s=(1,4/5,1/2)$ respectively). The inset in the right panel shows the $\e Q_0=0.6$ scalar amplitude from $V_{\rm EF}=500$ to $V_{\rm EF}=650$ after subtracting off the leading order power-law.}
    \label{fig:horizonfig}
\end{figure*}
\begin{figure*}[t]
    \centering
    \includegraphics[width=0.99\textwidth]{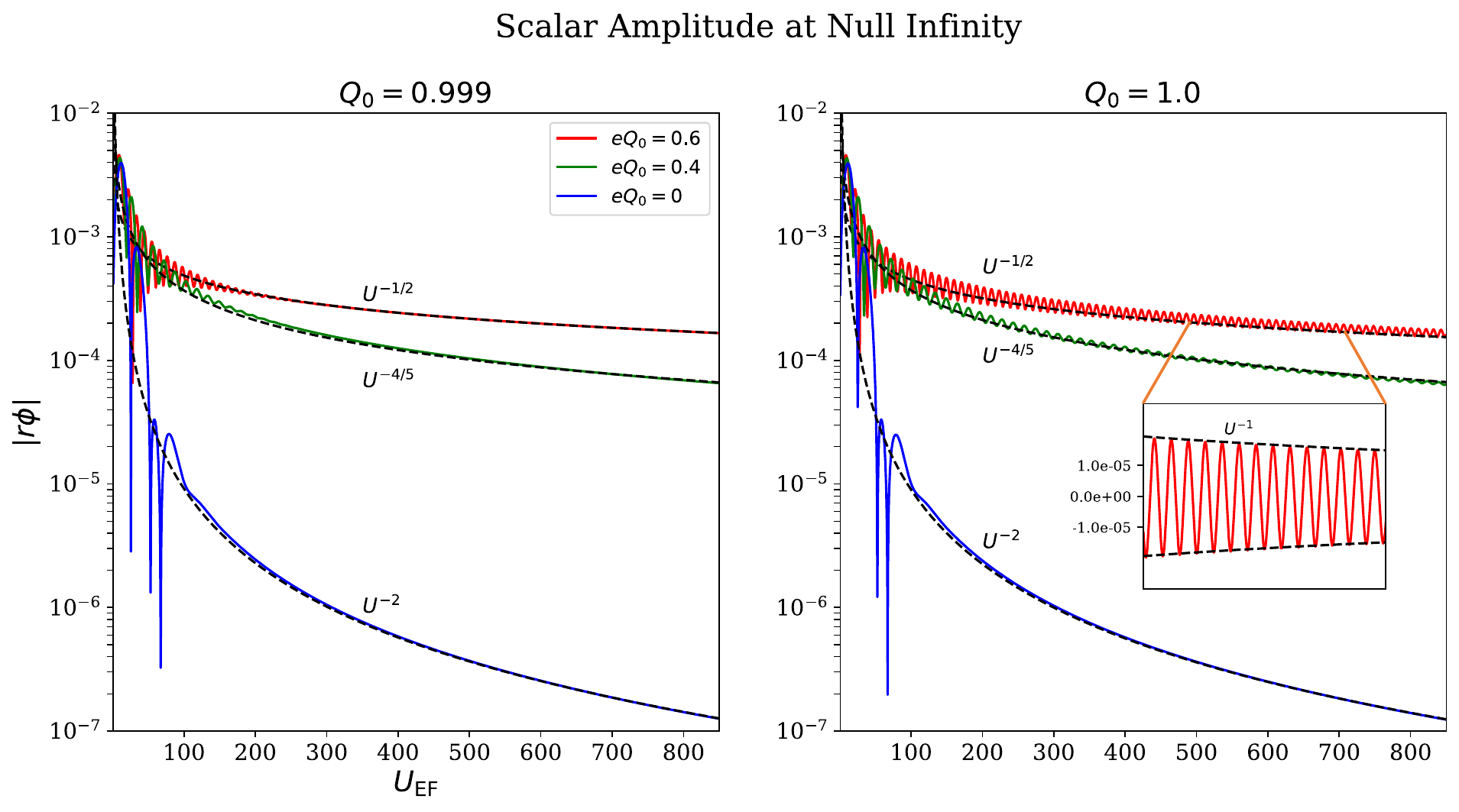}  
    \caption{Amplitude of the scalar field $|r\phi|$ at future null infinity $(v=\pi/2)$. The left panel shows results for the sub-extremal background, with the right panel showing results for the extremal background. Dashed lines indicate the approximate scalings for each power-law tail, fit by-eye, and guided by analytical expectations where known (see Table \ref{table:table1}, where $\e Q_0=(0,0.4,0.6)$ gives $s=(1,4/5,1/2)$ respectively). The inset in the right panel shows the $\e Q_0=0.6$ scalar amplitude from $V_{\rm EF}=500$ to $V_{\rm EF}=650$ after subtracting off the leading order power-law.}
    \label{fig:scrifig}
\end{figure*}

In Figures~\ref{fig:horizonfig} and \ref{fig:scrifig}, we plot the response of the scalar field on the future horizon and at future null infinity respectively. To enable comparison between the uncharged $(\e Q_0= 0)$ and charged $(\e Q_0\neq 0)$ perturbations, the plots are shown only for the $\tilde{\omega}=0$ perturbation, contrasting the different choices of $\e$ and $Q_0$ therein.

In each figure, we see strong oscillations, as well as power-law and/or exponential decay. As such, we can break down the response of the charged scalar field into three distinct segments: the prompt response, the quasi-normal modes (QNM's), and the power-law tail --- a triad familiar from black hole perturbation theory. The prompt response depends heavily on the specifics of the initial data, whereas the (complex) oscillation frequencies of the QNM's and the exponents of the power-law tail depend on the properties of the spacetime alone \cite{andersson_evolving_1997}. In particular, we see that the character of the QNM's and the power-law tail appear to depend discretely on: (a) whether the scalar field is charged and (b) whether the background spacetime is extremal. Below, we address both issues in detail, supplementing our numerical analysis with analytic arguments to better understand our novel results.

\subsubsection{Quasi-Normal Modes\label{sec:QNM}}
QNMs are damped sinusoids that dominate the response at intermediate times. The QNM's are evident for all cases in the left panel of Figure~\ref{fig:horizonfig}, in which the curves display a clear period of exponentially damped oscillations. However, we observe that there is a stark distinction between the nature of the QNM's in the charged $(\e Q_0> 0)$ cases and uncharged $(\e Q_0=0)$ case. While the uncharged scalar amplitude appears to reach zero periodically, the charged scalar does not; instead, it features exponential decay superposed with smaller oscillations.

We can attribute this distinction to electromagnetic gauge invariance. When $\e Q_0> 0$, the $U(1)$ gauge symmetry allows one to freely rotate the complex field by any local phase. So even though the solution to the charged wave equation on the future horizon can indeed be decomposed as a sum of damped sinusoids  \cite{leaver1985analytic,leaver_quasinormal_1990,hod1998late_2,hod_stability_2012,richartz_quasinormal_2014}:
\begin{align}
\label{eq:qnmdef}
   ( r\phi)|_{\mathcal{H}^+}&\sim \sum\limits_n A_ne^{i\omega_n V_{\rm EF}}
    \\\nonumber
   ( r\phi)|_{\mathcal{I}^+}&\sim \sum\limits_n A_ne^{i\omega_n U_{\rm EF}},
\end{align}
it is the {\em magnitude} of the {\em sum} of QNMs that is the gauge-invariant response. In this case, the real and imaginary components of the sum will generically not be in phase to exhibit zero crossings at the same moments in time. And unlike with a single real scalar field, one cannot perform the QNM analysis in the complex domain and then simply take the real part as the physical response.

Indeed, we find that the exponential decay of the uncharged  scalar fields $(\e Q_0=0)$ in Figure~\ref{fig:horizonfig} is very well fit by a \emph{single} QNM frequency: $\omega_{\rm uncharged}=0.133-0.096i$, which matches the fundamental mode computed analytically by Onozawa et al. \cite{Onozawa_extremal} at extremality. But for the charged scalar fields, we find that the horizon QNM's on the sub-extremal background are best described by a sum of \emph{two} frequencies $\omega_0$ and $\omega_1$, such that the scalar amplitude can be decomposed as:\begin{align}
\label{eq:qnm2}
    |r\phi|^2&=A_0^2e^{-2{\rm Im}[\omega_{0}]V_{\rm EF}}+A_1^2e^{-2{\rm Im}[\omega_{1}]V_{\rm EF}}\\&\nonumber+2A_0A_1e^{-{\rm Im}[\omega_{0}+\omega_{1}]V_{\rm EF}}\cos[{\rm Re}(\omega_{0}-\omega_{1})V_{\rm EF}+\varphi].
\end{align}
\begin{figure}[h]
    \centering
    \includegraphics[width=0.49\textwidth]{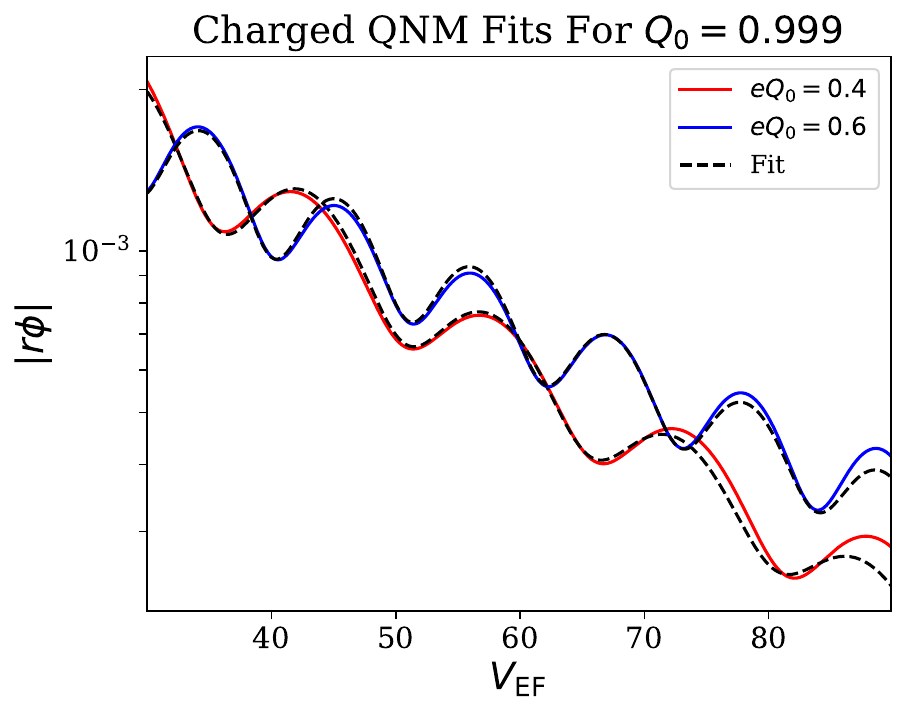}  
    \caption{Quasi-normal modes for the charged scalar field along the sub-extremal horizon. The fits assume that the response is dominated by two QNM's, for which the scalar amplitude should decompose as Eq.~\ref{eq:qnm2}. The values of the complex QNM frequencies are presented in Eq.~\ref{eq:qnmfit}, which are consistent with analytic predictions calculated using Leaver's method (Eq.~\ref{eq:qnmleaver}). The fits to the amplitude prefactors satisfy $|A_0/A_1|=3.5$ for $\e Q_0=0.4$ and $|A_0/A_1|=4.2$ for $\e Q_0=0.6$.}
    \label{fig:fitfig}
\end{figure}
We fit our numerical results to the above function, and we plot the results in Figure~\ref{fig:fitfig}. The fits to the fundamental ($\omega_0$) and overtone ($\omega_1$) frequencies are explicitly given by\begin{align}
\label{eq:qnmfit}
    {\rm Im}[\omega_{0,\rm fit}]&=\begin{cases}
    -0.034,&\e Q_0=0.6
    \\
    -0.026,&\e Q_0=0.4
    \end{cases}
    \\
    {\rm Im}[\omega_{1,\rm fit}]&=\begin{cases}
    -0.043,&\e Q_0=0.6
    \\
    -0.031,&\e Q_0=0.4
    \end{cases}
    \\
    {\rm Re}[\omega_{0,\rm fit}-\omega_{1,\rm fit}]&=\begin{cases}
    0.421,&\e Q_0=0.6
    \\
    0.576,&\e Q_0=0.4
    \end{cases}
\end{align}
To check that these frequencies are consistent with QNMs,
we use Leaver's method, as outlined in \cite{richartz_quasinormal_2014}, to compute the first two QNM's in Lorenz gauge : \begin{align}
\label{eq:qnmleaver}
    \omega_{0,\rm Leaver}&=\begin{cases}
    0.401-0.029i,& \e Q_0=0.4
    \\
        0.587-0.022i,& \e Q_0=0.6
    \end{cases}
    \\
    \omega_{1,\rm Leaver}&=\begin{cases}
        0.0012-0.040i,& \e Q_0=0.4
        \\
        0.0233-0.031i,& \e Q_0=0.6,
    \end{cases}
\end{align}
which all fall within 15\% of the numerical fits. The lack of exact agreement is not concerning because (a) Leaver's method starts to lose accuracy in the extremal limit\footnote{See Richartz \cite{richartz_quasinormal_2016} for a computation of charged scalar QNM's at extremality using the continued fraction method.}, (b) higher order overtones may be present, and (c) non-linearities from Maxwell's equations can introduce deviations from the simple damped sinusoid prescription. Thus, we conclude that at intermediate times, the oscillatory component of the charged scalar field we observe for the near-extremal $Q_0=0.999$ black hole is dominated by interference between two QNMs.

For both $\e Q_0=0.4$ and $\e Q_0=0.6$, the least damped QNM frequency has very small imaginary part $({\rm Im}\,\omega_0\ll 1)$. Each of these frequencies can be identified with a class of weakly damped modes that emerge at extremality. Hod \cite{hod_2010_v2,hod2012quasinormal} computed these modes analytically in Lorenz gauge as:\begin{align}\label{eq:hod_nzdm}
    \omega_{\rm NZD}|_{\rm Lorenz}&=\frac{\e Q_0}{r_+}-\frac{i\kappa_+}{2}+\mathcal{O}(\kappa_+^2).
\end{align}

Typically, these modes are referred to as ``Zero-Damped Modes" (ZDM's) \cite{hod_2010_v2,yang2013branching,Zimmerman_2016_v2}. But in this work, we will refer to them as ``Nearly Zero-Damped Modes" (NZDM's). We find that this is a more appropriate label because in practice this mode always decays; below extremality, $\omega_{\rm NZD}$ has a nonzero imaginary part and thus decays exponentially, and precisely at extremality, $\omega_{\rm NZD}$ becomes a point on a branch-cut in the Fourier-domain Green's function that gives rise to power-law decay \cite{casals_horizon_2016,zimmerman_horizon_2017}. So the NZDM \emph{always} decays, with the decay rate breaking from exponential to power-law at extremality. This phenomenon is discussed at length in the following subsection (and is related to the ideas discussed in \cite{casals_horizon_2016,richartz_synch_2}).

Extremal black holes also have damped QNMs \cite{Onozawa_extremal,Zimmerman_2016_v2,Yang_2013_v2,richartz_quasinormal_2016}. The fact that there is no prominent exponential decay at early times in the right panel of Figure~\ref{fig:horizonfig} implies our initial data is not exciting these QNMs with high enough amplitude to be visible above the power-law tail.
For the uncharged scalar ($\e Q_0=0$), a NZDM still exists, but its real part is identically zero~\cite{Zimmerman_2016_v2}. Similarly then, the fact that we do not see pure exponential decay at early times on the left panel of Figure~\ref{fig:horizonfig} implies our initial data is not exciting this NZDM with any appreciable amplitude.

It is worth noting that in our simulation output, the sub-extremal QNM's are not nearly as pronounced at null infinity as they are on the event horizon. This is likely because the power-law tails are shallower at null infinity, thus enabling them to mask the exponential decay of the short-lived QNM's. Indeed, the power-law exponents in Figures~\ref{fig:horizonfig} and \ref{fig:scrifig} depend sensitively on the values of $\e$ and $Q_0$, and we explore these differences below.

\subsubsection{Power-Law Tails\label{sec:powerlawtail}}
We show in this section that properties of the power-law tails we observe in our numerical simulations agree well with analytic predictions. In the linearized theory, these power-law tails are known to arise mathematically from a branch-cut in the Fourier-space Green's function that dominates the response at late times. 

Numerous authors have derived the power-law exponents for scalar fields in the regime where Maxwell's equations are ignored, the results of which are summarized in the first two rows of Table~\ref{table:table1}. We overplot these power laws as dashed lines in Figures~\ref{fig:horizonfig} and \ref{fig:scrifig}, finding excellent agreement between our numerical output and the analytic predictions (and we have confirmed, by resolutions studies, that we have converged to the quoted power-law exponents, at least to the level of accuracy of our ``by-eye'' fits). 

One important aspect of the power-law tails that can be seen in Figures~\ref{fig:horizonfig} and \ref{fig:scrifig} is the abrupt change in steepness between the tails of charged vs. uncharged scalar fields. This difference has been explicitly derived in analytic work. For example, on the sub-extremal future horizon, Price's law \cite{price_nonspherical_1972} predicts $V_{\rm EF}^{-3}$ decay for uncharged scalar fields. However, a field with nonzero coupling $(\e Q_0\neq 0)$ will decay more slowly due to its electromagnetic interaction with the black hole; as Hod \& Piran \cite{hod1998late_1,hod1998late_2,hod1998late_3} showed,\begin{align}
\label{eq:powercharge}
    |r\phi|_{\mathcal{H}^+}\sim V_{\rm EF}^{-2s} ,\qquad ( \e Q_0\neq 0,|Q_0|<1)
\end{align}
where \begin{align}
\label{eq:seq}
    s\equiv {\rm Re} \left(\frac{1}{2}+\sqrt{\frac{1}{4}-(\e Q_0)^2}\right)\in\left[\frac{1}{2},1\right]
\end{align}
is a parameter equal to $\frac{1}{2}$ for all $\e Q_0\geq \frac{1}{2}$. The discontinuous change between charged and uncharged fields is seemingly related to the phenomenon of electromagnetic gauge invariance. Namely, we do not expect $\e Q_0\to 0$ to be a continuous limit because the gauge symmetry is broken for $\e Q_0=0$ exactly (and then a complex scalar field does manifest two independent radiative ``degrees of freedom'').

Similarly, one can see from Figure~\ref{fig:horizonfig} that on the future horizon, there is a discontinuous change between the tails of the sub-extremal background vs. the tails of the extremal background. This discontinuity is related to the Aretakis instability, for which the power-laws break to shallower exponents\footnote{Note that the $V_{\rm EF}^{-2}$ scaling for the uncharged scalar in the right panel of Figure~\ref{fig:horizonfig} would break to an even shallower $V_{\rm EF}^{-1}$ if the outgoing initial data had support on the future horizon \cite{lucietti_horizon_2013,angelopoulos2020late}.} when $Q_0=1$ \cite{aretakis2011stability1,aretakis2011stability2,lucietti_horizon_2013,angelopoulos2020late}. Indeed, when $\e Q_0\neq 0$, the power-law exponents on extremal backgrounds are exactly half of their value on sub-extremal backgrounds, as shown analytically by Zimmerman \cite{zimmerman_horizon_2017}. The Aretakis instability, however, does not affect the behavior of the scalar field at null infinity, which explains the agreement between the left and right panels of Figure~\ref{fig:scrifig}.

We note that the analytic predictions for the power-law tails were all computed in the linearized theory (i.e. solving \emph{only} the wave equation on a static background metric and electromagnetic potential, thus fixing $A_\mu$ in Eq.~\ref{eq:waveeqn}). It is reassuring to see that when solving the non-linear coupled system of charged-scalar/Maxwell equations, as we have in our numerical solutions (though still on a fixed metric background), the leading order power-law scalings do not change. This finding makes sense for our choice of initial data, which was intentionally constructed to satisfy Eq.~\ref{eq:qestimate} and hence keep the non-linearities of SED small throughout evolution. Unlike what we would expect from Einstein's equations \cite{murata_what_2013}, the non-linearities of Maxwell's equations remain in check even at extremality. Though as we will see in \S\ref{sec:chargedensitysec}, the dynamical Maxwell field generates an instability of its own on the extremal horizon.

Even so, there is one feature of the power-law tails in Figures~\ref{fig:horizonfig} and \ref{fig:scrifig} that is not discussed in past analytic work. In both figures, the charged ($\e Q_0\neq 0)$ power-law tails on the extremal background are superposed with oscillations, and these oscillations appear to decay slowly. As we detail below, we can attribute this phenomenon once again to the limiting behavior of the NZDMs.

Working with linearized scalar electrodynamics in Lorenz gauge, Hod \& Piran \cite{hod1998late_2} derived an explicit form of the Fourier-space Green's function $\tilde{G}_L(\omega)$, which represents the response of the charged scalar field to a delta function source. Consistent with prior results, Hod \& Piran showed that on a sub-extremal background, $\tilde{G}_L$ has a branch cut along the negative imaginary axis (${\rm Re}\,\omega=0$), and the Fourier integral around this branch cut ultimately produces the late-time power law tail. Specifically, they showed (their Eq.~30) that the Fourier-space Green's function scales as $\tilde{G}_L\sim \omega^{2s-1}$
near the origin, where the Green's function has the most support. 

Zimmerman \cite{zimmerman_horizon_2017} then showed that at extremality this branch cut persists, but a \emph{second} branch cut appears at ${\rm Re}\,\omega=\omega_{\rm NZD}$, whose leading order contribution to the Fourier integral scales as $(\omega-\omega_{\rm NZD})^{s-1}$. Thus, the time-domain Green's function on an extremal background will scale as\begin{align}
\label{eq:ftgreen}
    G_L(V_{\rm EF})= {\rm FT}[\tilde{G}_L(\omega)]\sim {\rm FT}[\omega^{2s-1}]+{\rm FT}[(\omega-\omega_{\rm NZD})^{s-1}],
\end{align}
where ``FT" denotes Fourier Transform.
Zimmerman evaluates only the second term, arguing that it produces the leading order power-law decay in the time domain. However, we claim that the first term is dynamically important too, and it ultimately produces the late-time oscillations seen in our numerical simulations.

To see this, we employ the Fourier shift theorem: that a shift in the Fourier domain is equivalent to multiplication by a phase in the time domain. Using this theorem, the Fourier Transforms in Eq.~\ref{eq:ftgreen} can be straightforwardly evaluated as\begin{align}
    G_L(V_{\rm EF})\sim V_{\rm EF}^{-2s}+e^{i\omega_{\rm NZD} V_{\rm EF}}V_{\rm EF}^{-s}.
\end{align}
The presence of the phase $e^{i\omega_{\rm NZD} V_{\rm EF}}$ indicates that interference will occur between these two terms. Taking the modulus of the above expression, the gauge-invariant amplitude of the scalar field on the extremal horizon will scale as
\begin{align}
    \label{eq:oscpower1}
    |r\phi|_{\mathcal{H}^+}&\sim|G_L|\nonumber\\
    &\sim A_V V_{\rm EF}^{-s}+B_V V_{\rm EF}^{-2s}\left[1+\epsilon_V \cos\left(\e V_{\rm EF}+\varphi_V\right)\right],
\end{align}
where $A_V,B_V,\varphi_V$ and $\epsilon_V$ are constants, and $\omega_{\rm NZD}=\e$ is the limiting value of the NZDM frequency (Eq.~\ref{eq:hod_nzdm}) at extremality. So indeed, ``interference'' between the branch point at $\omega=0$ and the branch point at $\omega= \e$ produces oscillations in the late-time response of the scalar field as shown in the right panel of Figure~\ref{fig:horizonfig}.

By analogous logic, we expect a similar oscillation to modulate the decay at null infinity: \begin{align}
\label{eq:oscpowerscri}
    |r\phi|_{\mathcal{I}^+}\sim A_UU_{\rm EF}^{-s}+B_U U_{\rm EF}^{-2s}[1+\epsilon_U \cos(\e U_{\rm EF}+\varphi_U)]
\end{align}
with constants $A_U,B_U,\varphi_U$ and $\epsilon_U$. 
This interference pattern is evident in the right panel of Figure~\ref{fig:scrifig}. 

\begin{figure}[h]
    \centering
    \includegraphics[width=0.49\textwidth]{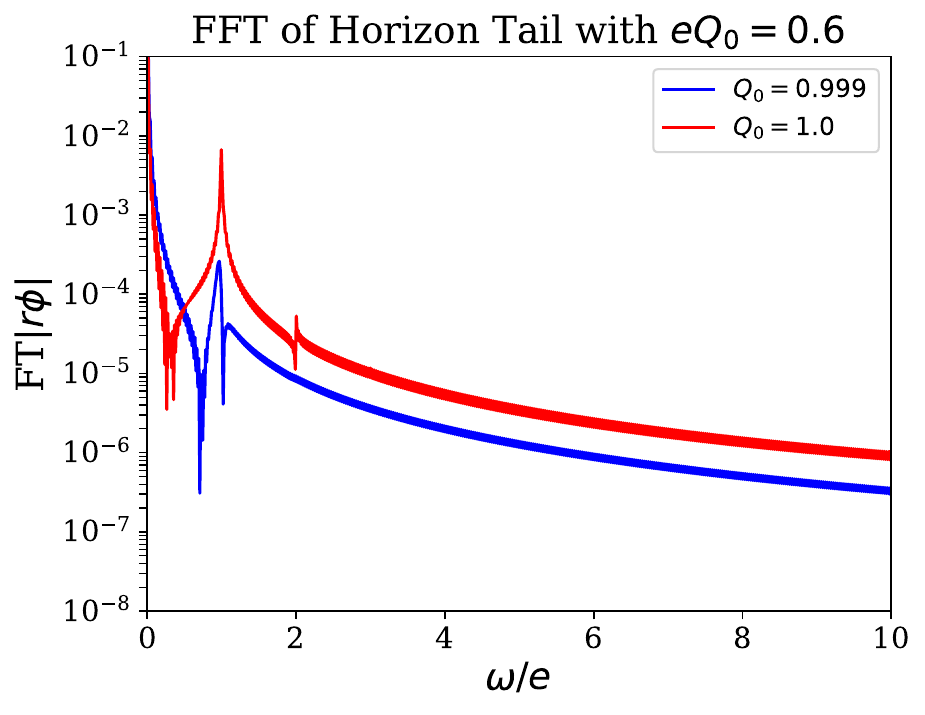}  
    \caption{Fourier Transform of horizon power-law tails (sampled from $V_{\rm EF}=500$ to $V_{\rm EF}=900$) for an extremal and sub-extremal spacetime. Both FFT's take the form of power-laws, with the extremal case showing a significant peak around the limiting frequency of the NZDM. The ``scatter'' in each curve manifesting as seemingly thick lines in the figure is a numerical artifact due to finite sampling in the Fourier domain.} 
    \label{fig:fftfig}
\end{figure}

More than implying oscillations, Eqs.~\ref{eq:oscpower1} and \ref{eq:oscpowerscri} predict the oscillations decay as power-laws with rates $V_{\rm EF}^{-2s}$ and $U_{\rm EF}^{-2s}$ respectively, and oscillate with frequency  $\omega_{\rm NZD}$ measured in the appropriate EF coordinate. For the decay rates, the insets on the right panels of Figures ~\ref{fig:horizonfig} and \ref{fig:scrifig} illustrate that we do see this expected behavior. To verify the oscillation frequency prediction, we take a Fast Fourier Transform (FFT) of the scalar field amplitude at late time (to remove influence of the early time transients). This is shown in Figure~\ref{fig:fftfig} on the horizon for the $\e Q_0=0.6$ extremal case, and to compare we also show the subextremal case.

In the extremal limit, a clear peak in the FFT emerges near $\omega=\e$. Specifically, we measure the peak of the FFT to be located at\begin{align}
\label{eq:ZDMpeak}
    \frac{\omega_{\rm Peak}}{\e}&=0.995,
\end{align}
which is consistent with $\frac{\omega_{\rm peak}}{\e}=1$ since we have a Fourier-domain resolution of $\frac{\Delta\omega}{\e}=0.013$ for this particular simulation. 

A smaller, smoother peak near $\omega=\e$ is also visible for the sub-extremal (blue) FFT in Figure~\ref{fig:fftfig}. This is consistent with the notion that the Fourier-space Green's function is large --- but still formally finite --- at the NZDM frequency for near-extremal black holes; the branch cut emerges only precisely at extremality.

In sum, the NZDMs in the extremal limit most pronouncedly affect the late-time dynamics of the charged scalar field by making the power-law tail shallower, and give a sub-leading oscillation on top of this tail at a frequency $\omega=\e$. For the uncharged scalar, the NZDMs merge with the existing branch cut at $\omega=0$, altering the power-law tail steepness but failing to produce any sort of oscillatory interference pattern. 

Interestingly, the tail interference appears to be a unique feature of charged scalar fields incident on extremal Reissner-Nordstr\"om black holes: we do \emph{not} expect the same phenomenon to occur for real fields incident on extremal Kerr black holes. In the latter case, the Fourier-space Green's function similarly contains two branch points (one at $\omega=0$ and one at $\omega=m/2$, where $m$ is the azimuthal mode number) \cite{glampedakis2001late,casals_horizon_2016}. However, the time-domain response will be given by taking the real part of the Green's function, as opposed to the modulus. So while some transient beating between different modes may occur, we expect a single dominant mode to occupy the response at late times.

Finally, we remark that while Figures~\ref{fig:horizonfig} and \ref{fig:scrifig} show our numerical results only for the simulation with neutral initial data $(\tilde{\omega}=0)$, every result we have presented in this section still holds for nonzero values of $\tilde{\omega}$. Indeed, we explicitly verified that the exponential decay rate, power-law scalings, and NZDM-influenced oscillations show up in the same manner for $\tilde{\omega}\in\{\e/2,\e,3\e/2\}$. Also, note that the scalar field with initial data $\tilde\omega=0$ does, upon evolution, exhibit charge separation (though the net charge of the perturbation remains zero), and therefore even this case features non-trivial charge density dynamics on the future horizon.

While $\tilde{\omega}$ appears to have a negligible effect on the dynamics of the scalar field, we discuss below how it affects the behavior of the enclosed charge $Q$.

\subsection{Asymptotic Behavior of Enclosed Charge \label{sec:Qevolve}}
For the class of initial data we explored, we find that $Q(U,V)$ exhibits a short burst of transient deviation from $Q_0$, before settling down to some final value $Q_f$. Without incorporating Einstein's equations, $Q_f$ will not be reflected in the underlying spacetime geometry. But because our initial data has small amplitude, $Q_f$ will always be close to $Q_0$, ensuring that the resulting disconnect remains small. As an additional consequence, the non-linearities of SED --- while fully incorporated into our evolution scheme --- remain small too.

\begin{figure}[h]
    \centering
    \includegraphics[width=0.49\textwidth]{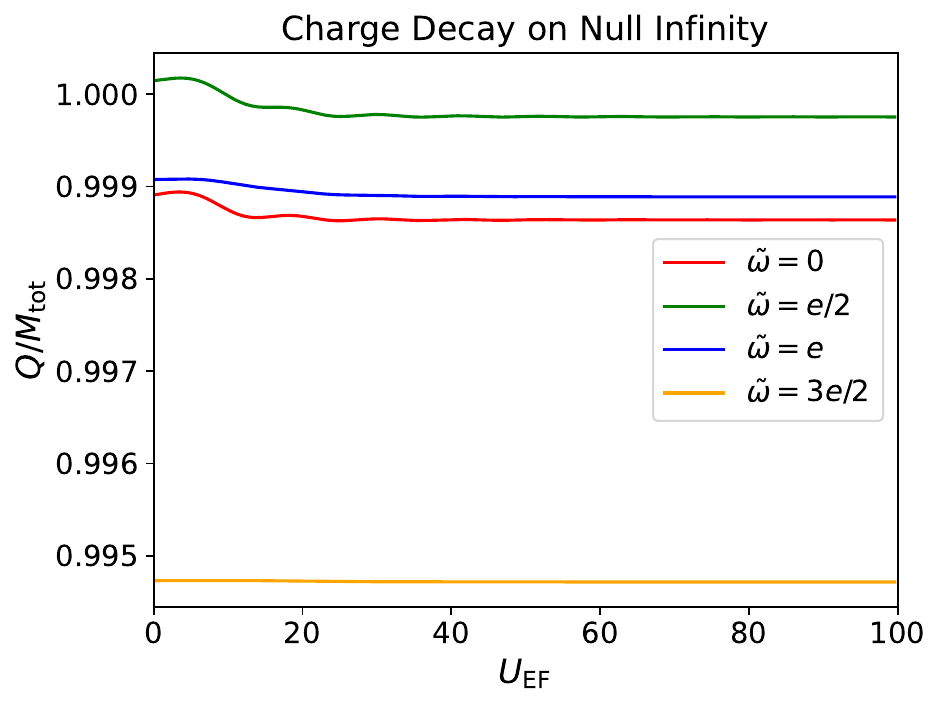}  
    \caption{Charge-to-mass ratios plotted along null infinity. This plot is shown specifically for the case with $Q_0=1.0$ and $\e Q_0=0.6$, thus enabling direct comparison to Figure~\ref{fig:initdatafig}.} 
    \label{fig:chargescri}
\end{figure}

The particular dynamics of $Q(U,V)$ do depend on the initial data frequency $\tilde{\omega}$. To anticipate how this may cause evolution away from extremality in our planned follow-up study with metric back-reaction, we investigate this dynamics at future null infinity using the ratio $Q/M_{\rm tot}$, with $M_{\rm tot}$ defined as follows. First, we use a similar expression to Eq.~\ref{eq:Macc} (for $M_{\rm acc}$) to compute the mass flux through null infinity:\begin{align}
    \frac{dM_{\rm tot}}{dU_{\rm EF}}\bigg|_{\mathcal{I}^+}&=\lim_{r\to\infty}-4\pi r^2\mathcal{T}_{U_{\rm EF}U_{\rm EF}}\\&=-2\left[\left(\frac{Q_{,U_{\rm EF}}}{8\pi \e P}\right)^2+(P_{,U_{\rm EF}})^2\right]\bigg|_{\mathcal{I}^+}.
\end{align}
We then integrate this quantity to give $M_{\rm tot}(U_{\rm EF})$, with $M_{\rm tot}({U_{\rm EF}}=0)=1+M_{\rm acc}(V_{\rm EF}=\infty)$. In Figure~\ref{fig:chargescri}, we plot $Q/M_{\rm tot}$ for a subset of our numerical simulations.

As we can see in the cases studied here, the charge-to-mass ratio always ends below where it started, even for the initial data with $\tilde\omega=\e/2$ and $(Q_0/M_{\rm tot})|_{\mathcal{N}_B}>1$ (see Fig.~\ref{fig:initdatafig}). This is not surprising given that SED allows charge separation in the scalar field, so one would expect a positively charged black hole ($Q_0>0$) to preferentially accrete negative charge, while repelling positive charge density to infinity. Consequently, one would expect that if metric backreaction were included, the spacetime would evolve away from extremality, mitigating the Aretakis instabilities discussed in the following section. 

This may be generic, but of course we cannot do more than speculate given our limited set of initial data, and given that we do not evolve the Einstein equations. Even so, analogous to what MRT found for the uncharged scalar field case~\cite{murata_what_2013}, we would expect to be able to construct fine-tuned initial data that would at least be able to maintain extremality at late times in the presence of backreaction. 

\subsection{Horizon Instabilities}
Now that we have covered the decay of the scalar amplitude and enclosed charge, we will turn to the behavior of their transverse derivatives: $\p_r|r\phi|$ and $\p_r Q$. As we will see, both derivatives do not decay on the extremal horizon, consistent with the expected behavior of the Aretakis instability and its enhancement when the field is charged. 

\subsubsection{Energy Density}
As mentioned, our motivation to study transverse derivatives along the extremal event horizon comes from the well-known Aretakis instability. This instability was originally formulated for uncharged fields \cite{aretakis2011stability1,aretakis2011stability2} and concerned the non-decay of transverse derivatives $\p_r(r\phi)$. 

To see where the instability comes from, consider the uncharged wave equation for a real scalar field in double-null coordinates:\begin{align}
  (r\phi)_{,UV}=(r\phi)g,\qquad (\e=0).
\end{align}
On the event horizon, the spacetime source term in the wave equation (Eq.~\ref{eq:gdefn}) becomes \begin{align}
    g|_{r=r_+}=-\frac{\kappa_+f}{r_+},
\end{align}
which vanishes at extremality and hence implies that the quantity 
\begin{align}\label{H0_def}
H_0\equiv (r\phi)_{,U}|_{r=r_+}\propto (r\phi)_{,r}|_{r=r_+}
\end{align} 
will be constant (an analogous explanation is given in \cite{lucietti_horizon_2013}).
This non-decay of transverse derivatives conflicts with our notion of linear stability, which is a known property of sub-extremal RN black holes \cite{giorgi2020linear}. Indeed, one can show that conservation of $H_0$ implies that higher-order transverse derivatives (e.g. $(r\phi)_{,rr}$) will grow without bound along the extremal event horizon \cite{aretakis2011stability1,lucietti_horizon_2013}. 

Physically, the radial derivative of a scalar field is a kinetic term and therefore tracks a local energy density. In terms of this physical interpretation, the Aretakis instability makes intuitive sense: extremal black holes have no horizon redshift $(\kappa_+=0)$, and the horizon is only a marginally trapped surface, implying outgoing energy density on and near the horizon on the inside will not decay \cite{lucietti_horizon_2013}. 

We can extend this notion to scalar electrodynamics by computing the net energy density of the charged scalar and electromagnetic fields. Using Eq.~\ref{eq:tdef}, the energy density $\rho_{E,\rm net}$ on the event horizon as measured by a timelike observer with four-velocity $u^\mu$, free falling into the black hole starting from rest at infinity, is given by\begin{align}
\label{eq:rhoE_net}
    \rho_{E,\rm net}|_{r=r_+}&=u_\mu u_\nu T^{\mu\nu}|_{r=r_+}
    \\&=\nonumber
   \frac{2}{r_+^2}\left[\left(\frac{Q_{,r}}{8\pi \e P}\right)^2+r_+^2\left(\frac{P}{r}\right)_{,r}^2\right]\\\nonumber&+\frac{1}{2r_+^2}\left[\left(\frac{Q_{,V_{\rm EF}}}{8\pi \e P}\right)^2+P_{,V_{\rm EF}}^2\right]+\frac{Q^2}{16\pi r_+^4}.
\end{align}
The final term $(Q^2/16\pi r_+^4)$ simply represents the Coulombic energy stored in the black hole's electric field; this term is not relevant for our purposes, so we will ignore it going forward, focusing on the following quantity
\begin{align}
\label{eq:rhoE}
    \rho_{E}&\equiv \rho_{E,\rm net}|_{r=r_+} - \frac{Q^2}{16\pi r_+^4}.
\end{align}

\begin{figure*}[t]
    \centering
    \includegraphics[width=0.99\textwidth]{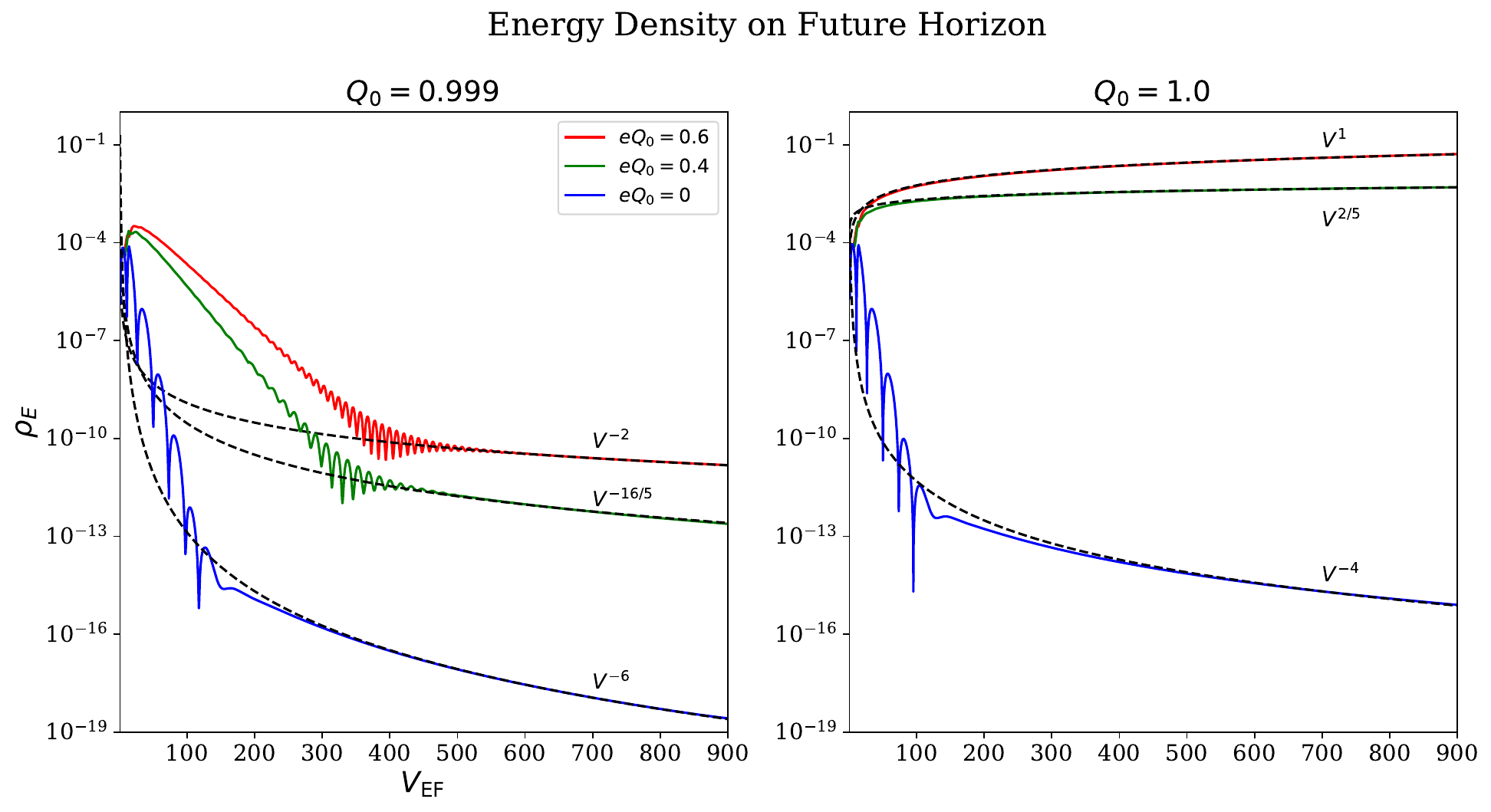}  
    \caption{Energy density on the future horizon, excluding the Coulomb part of the electric field, as measured by a time-like observer free falling from rest at infinity (Eq.~\ref{eq:rhoE}). The left panel shows results for the sub-extremal background, while the right panel shows results for the extremal background. Dashed lines indicate by-eye power-law fits to the late time behavior of each case, guided by the analytical expectations (see Eqs.~\ref{eq:rho_Q1}-\ref{eq:rho_Qlt1}, where $\e Q_0=(0,0.4,0.6)$ gives $s=(1,4/5,1/2)$ respectively).}
    \label{fig:energyfig}
\end{figure*}

Using the same finite difference stencils described in Appendix~\ref{app:numerics}, we compute $\rho_E|_{r=r_+}$ in the post-processing of each of our numerical simulations. In Figure~\ref{fig:energyfig}, we plot the results for the simulations with $\tilde{\omega}=0$, illustrating the role of $\e$ and $Q_0$ in controlling $\rho_E$. 

We see that, as expected, each of the sub-extremal energy densities (left panel of Figure~\ref{fig:energyfig}) does indeed decay. In fact, the energy density for the uncharged scalar $(\e Q_0=0)$ decays on the extremal background too. This is because these simulations are all initialized with \emph{ingoing} data, which satisfy $H_0=\p_r(r\phi)|_{r=r_+}=0$ by construction (see Table~\ref{table:table1}). 
However, the energy densities of the charged $(\e Q_0\neq 0)$ scalars on the extremal background all grow as power-laws in time. So contrary to the uncharged Aretakis instability, the charged instability of the first transverse derivative can be triggered for purely ingoing initial data. 

An explanation for this distinction follows naturally from the gauge-invariant evolution equations (Eqs.~\ref{eq:peq}-\ref{eq:qeq}). On the extremal event horizon, the spacetime source term for the scalar wave equation again disappears $(g=0)$. However, a Maxwell coupling term remains, and the evolution equation for $Q$ retains all its source terms  as well ($f_{\rm MRT}\neq0$ in on the horizon, see Eq.~\ref{eq:metriclimit}). That is, Eqs.~\ref{eq:peq}-\ref{eq:qeq} become
\begin{align}
    P_{,UV}|_{r=r_+}&=-\frac{Q_{,U}Q_{,V}}{64\pi^2\e^2P^3}\Bigg|_{r=r_+},\label{eq:peq_h}
    \\
    Q_{,UV}|_{r=r_+}&=\left[\frac{P_{,U}Q_{,V}}{P}+\frac{P_{,V}Q_{,U}}{P}-\frac{4\pi \e^2fP^2Q}{r^2}\right]_{r=r_+} .\label{eq:qeq_h}
\end{align}
Suppose then we begin with initial data that is the analogue of $H_0=0$, namely $P=0,P_{,U}=0$ and hence $Q_{,U}=0$ at $r=r_+$. Integrating in $V$ along the horizon, $P_{,U}$ and $Q_{,U}$ will remain $0$ as long as $P=0$ (and note that when $P=0$ the seemingly singular terms in Eqs.~\ref{eq:peq_h} and ~\ref{eq:qeq_h} evaluate to zero using Eqs.~\ref{eq:maxwellcons1},~\ref{eq:maxwellcons2} and \ref{eq:currentdef2}). However, as soon as incoming radiation ($P\neq0, P_{,V}\neq0, Q_{,V}\neq0$) reaches the horizon, the third (non-linear) source term on the RHS of Eq.~\ref{eq:qeq_h} will cause $Q_{,U}$ to evolve away from zero, and hence by Eq.~\ref{eq:peq_h} so will $P_{,U}$. Thus for a charged scalar field, $H_0$ is {\em not} conserved on an extremal horizon.

Indeed, the source term on the righthand side of the above equation represents the electromagnetic interaction between the scalar field and the black hole, in this case serving to increase the local energy density. This accounts for the growth of the energy density shown in the right panel of Figure~\ref{fig:energyfig}.

Zimmerman \cite{zimmerman_horizon_2017} explicitly calculated the expected growth rates from a Green's function formalism in the linearized theory. Relying on the existence of the branch point at $\omega=\omega_{\rm NZD}$ (i.e. Eq.~\ref{eq:ftgreen}), his results imply\begin{align}
\label{eq:Pgrow}
    P_{,r}|_{r=r_+}\sim V_{\rm EF}^{1-s},\quad (\e Q_0\neq 0, Q_0=1)
\end{align}
where $s$ is defined in Eq.~\ref{eq:seq}. Assuming that the $(P_{,r})^2$ term is the dominant term in the expression for $\rho_E$, this in turn suggests that \begin{align}\label{eq:rho_Q1}
    \rho_E|_{r=r_+}\sim V_{\rm EF}^{2-2s},\quad (\e Q_0\neq 0, Q_0=1).
\end{align}
In Figure~\ref{fig:energyfig}, we overplot this prediction, finding again excellent agreement between our numerical results and the analytics of Zimmerman \cite{zimmerman_horizon_2017}. 

In the left panel of Figure~\ref{fig:energyfig}, we also overplot the analytically predicted energy decay for the sub-extremal background, which follows from the results of Hod \& Piran \cite{hod1998late_2}:\begin{align}\label{eq:rho_Qlt1}
    \rho_E|_{r=r_+}\sim V_{\rm EF}^{-4s}, \quad (\e Q_0\neq 0, Q_0<1).
\end{align}
This result also appears to give excellent fits to the numerical data (and as with our results in the previous subsection, we have confirmed convergence to these expected power-law exponents to the level of accuracy of our ``by-eye'' fits).

In sum, our numerical simulations support the claim that
\begin{align}
    \lim_{V_{\rm EF}\to \infty}\rho_E\to
    \begin{cases}
    \infty ,&|Q_0|=1,\,\e Q_0 \neq 0
    \\
    {\rm const},&|Q_0|=1,\,\e Q_0=0
    \\
    0,&{\rm otherwise},
    \end{cases}
\end{align}
when the wave equation and Maxwell's equations are evolved together (and again, for the uncharged case we have only evolved $H_0=0$ initial data, for which the constant in the middle row above is $0$). The power-law decay exponents of the horizon energy densities are summarized in the third row of Table~\ref{table:table1}.

While these results indicate that the growth of $\rho_E$ on the extremal event horizon comes predominantly from the transverse derivative of the scalar field amplitude, it is also worth examining the transverse derivative of the enclosed charge --- $Q_{,r}$ --- as this term also contributes to the expression in Eq.~\ref{eq:rhoE_net}. To isolate the effects of $Q_{,r}$, we explore the charge density on the extremal horizon.

\subsubsection{Charge Density}
\label{sec:chargedensitysec}
Until now, all analyses of scalar electrodynamics on fixed RN backgrounds (e.g. \cite{hod1998late_1,hod1998late_2,hod1998late_3,hod_2010_v2,hod_stability_2012,richartz_quasinormal_2014,zimmerman_horizon_2017}) have neglected the self-force of the matter and treated the electric field as a time-independent function sourced only by the central black hole. While this approximation captures the correct dynamics of the scalar field to linear order, one might wonder if there are any potential non-linear effects that might influence the evolution of the charge density coming from Maxwell's equations near the extremal horizon. 

However, in this work we have limited the amplitude of our initial data to give only small perturbations of the charge (to avoid issues interpreting a significant deviation of $Q$ relative to its background value). Moreover, in spherical symmetry, the dynamics of $Q$ are completely driven by that of the scalar field. Therefore, we do not expect any qualitative deviations in our numerical results of the late-time behavior compared to that found using the linear analyses discussed above. In this section, we show how the Aretakis instability manifests in the charge density on the horizon, and that the late-time behavior is indeed consistent with its manifestation in the energy density of the scalar field.  

\begin{figure*}[t]
    \centering
    \includegraphics[width=0.99\textwidth]{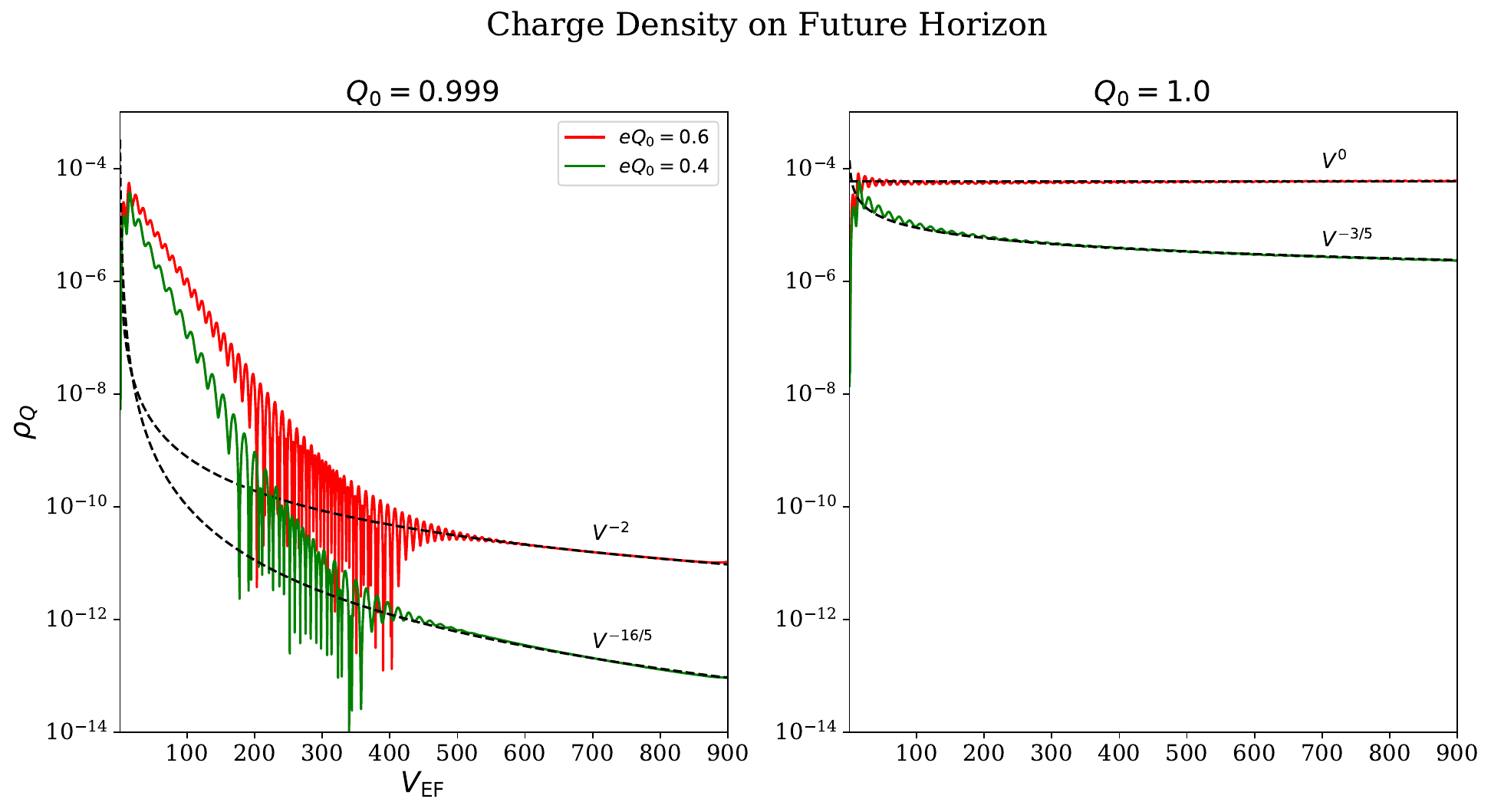}  
    \caption{Charge density on the future horizon, as measured by a time-like observer free falling from rest at infinity. The left panel shows results for the sub-extremal background, with the right panel showing results for the extremal background. Dashed lines indicate the approximate scalings for each power-law tail (see Eqs.~\ref{eq:rhoQ_pl}, where $\e Q_0=(0.4,0.6)$ gives $s=(4/5,1/2)$ respectively).} 
    \label{fig:chargefig}
\end{figure*}

As measured by the same family of timelike free-falling observers introduced in the previous section, the charge density on the horizon is given by\begin{align}
    \rho_Q|_{r=r_+}&=-u^\mu J_\mu|_{r=r_+}=\frac{Q_{,r}+Q_{,V_{\rm EF}}/2}{4\pi r_+^2}.
\end{align} 
As we did with the energy density, we use finite differences to compute $\rho_Q|_{r=r_+}$ in the post-processing of our numerical simulations. Results for the $\tilde{\omega}=0$ runs are shown in Figure~\ref{fig:chargefig}.

Note that the charge density on the horizon is always positive at late times. At first glance, this seems counter-intuitive, as one would expect that negative charge accretes onto the black hole while positive charge is repelled to infinity. In reality, this {\em does} happen, but charge separation continues on the horizon itself, leading to positive charge density that cannot escape to infinity and instead accumulates on the horizon.

From Maxwell's equations (Eqs.~\ref{eq:maxwellcons1}-\ref{eq:maxwellcons2}), when gradients of the scalar field dominate (as they do at late times when the Aretakis instability is active), the electromagnetic current (Eq.~\ref{eq:currentdef2}) is dominated by terms like $\phi\nabla_a\phi$. Using the results from the previous section, this suggests the power-law dynamics on the horizon should scale as $\rho_Q \propto \sqrt{\rho_E} P$, giving (for $\e Q_0\neq0$)
\begin{align}
    \rho_Q|_{r=r_+}&\sim\begin{cases}
    V_{\rm EF}^{-4s},  &|Q_0|<1 \\
    V_{\rm EF}^{1-2s}, &|Q_0|=1. \label{eq:rhoQ_pl}
    \end{cases}
\end{align}
Indeed, these scalings for $Q_{,r}$, combined with the scalings for $P_{,r}$ derived in the previous subsection, can easily be shown to form an asymptotic solution to the gauge-invariant system of PDE's (Eqs.~\ref{eq:peq}-\ref{eq:qeq}) when evaluated on the extremal horizon.
And as can be seen in Figure \ref{fig:chargefig}, these scalings match the numerical results well. 

For strongly coupled matter with $|\e Q_0|\geq 1/2$, we have $s=1/2$, yielding a transverse derivative of the charge that approaches a constant on the extremal event horizon. This represents an instability of the charge density. Similar to the original formulation of the uncharged Aretakis instability, the addition of SED has given rise to a new asymptotic constant $\rho_Q$ on the extremal horizon (though the analogous Aretakis quantity $H_0$ is exactly a constant). We speculate on the physical origin of this new aspect of the instability in the following section.

In sum, our numerical solutions support the claim that\begin{align}
    \lim_{V_{\rm EF}\to \infty}\rho_Q|_{r=r_+}&\to\begin{cases}
    {\rm const},&|\e Q_0|\geq 1/2, |Q_0|=1
    \\
   0,&{\rm otherwise.}
    \end{cases}
\end{align}
when the wave equation and Maxwell's equations are evolved together. The power-law decay exponents of the horizon charge densities are summarized in the fourth row of Table~\ref{table:table1}.

It will be interesting to examine what happens in the regime where the initial data amplitude is large and the true non-linearities of SED are exposed, but this can be done self-consistently only when metric backreaction is incorporated too.

\section{Results - Monochromatic Initial Data\label{sec:superradiance}}
The unique behavior of scalar electrodynamics on the  extremal RN horizon can (in many cases) be attributed to the existence of the NZDM. Indeed, we have seen in the previous section that the NZDM gives rise to shallower power-law tails at extremality, as well as oscillations in the late time response of the scalar amplitude. Furthermore, Zimmerman's \cite{zimmerman_horizon_2017} analytic derivation of the charged Aretakis instability relies explicitly on the existence of a branch point in the Lorenz-gauge Green's function at $\omega=\omega_{\rm NZD}$. Richartz, Herdeiro, \& Berti \cite{richartz_synch_2} then expanded upon this analysis, demonstrating that the NZDM is not a normal mode but rather a scattering mode that can enhance the horizon instabilities.

However, the unique role of the NZDM is difficult to tease out in our numerical simulations for which the initial data is compactly supported. With compact support, the Lorenz-gauge Fourier decomposition of the initial data contains infinitely many modes, each of whose amplitude is formally zero. So to isolate the role of the NZDM in enhancing the Aretakis instability, we perform numerical simulations with \emph{monochromatic} initial data consisting of a single dominant mode.

Specifically, we repeat the numerical procedure of the previous section, but we take the scalar envelope $\mathcal{A}(V)$ to consist of a smoothed step function:
\begin{align}
    \mathcal{A}(V)&=\begin{cases}
    0,&V_{\rm EF}\leq 0
    \\
    \frac{0.001}{1+\exp\left[\frac{15}{4}\left(\frac{1}{V_{\rm EF}}+\frac{1}{V_{\rm EF}-15}\right)\right]},&0\leq V_{\rm EF}\leq 15
    \\
    0.001 ,&V_{\rm EF}\geq 15.
    \end{cases}
\end{align}
This function smoothly rises from 0 to $0.001$ and then remains constant for all $V_{\rm EF}\geq 15$. This transition will produce a non-monochromatic transient, but the late-time dynamics will be driven by a monochromatic incoming wave. Note that we have reduced the maximum amplitude of the scalar field by a factor of 10 compared to the compactly supported case; this is necessary to keep the the net charge accumulated on $\mathcal{N}_B$ small, hence ensuring that the disconnect between $Q_f$ and $Q_0$ remains in check. Indeed, as one can see from Eq.~\ref{eq:qestimate}, the reduction in amplitude is necessary to push the non-linearities of SED past $V_{\rm EF}\sim 1000M$.

We then run simulations with the same set of parameters that were used in the previous sections (Eqs.~\ref{eq:qpar}-\ref{eq:omegapar}). However, to isolate the role of the NZDM frequency, we choose a narrower band of values for $\tilde{\omega}$:\begin{align}
    \tilde{\omega}\in\{0.9\e,0.99\e,\e,1.01\e,1.1\e\},
\end{align}
with $\tilde{\omega}=\e$ corresponding to the NZDM at extremality. 

\begin{figure*}[t]
    \centering
    \includegraphics[width=0.8\textwidth]{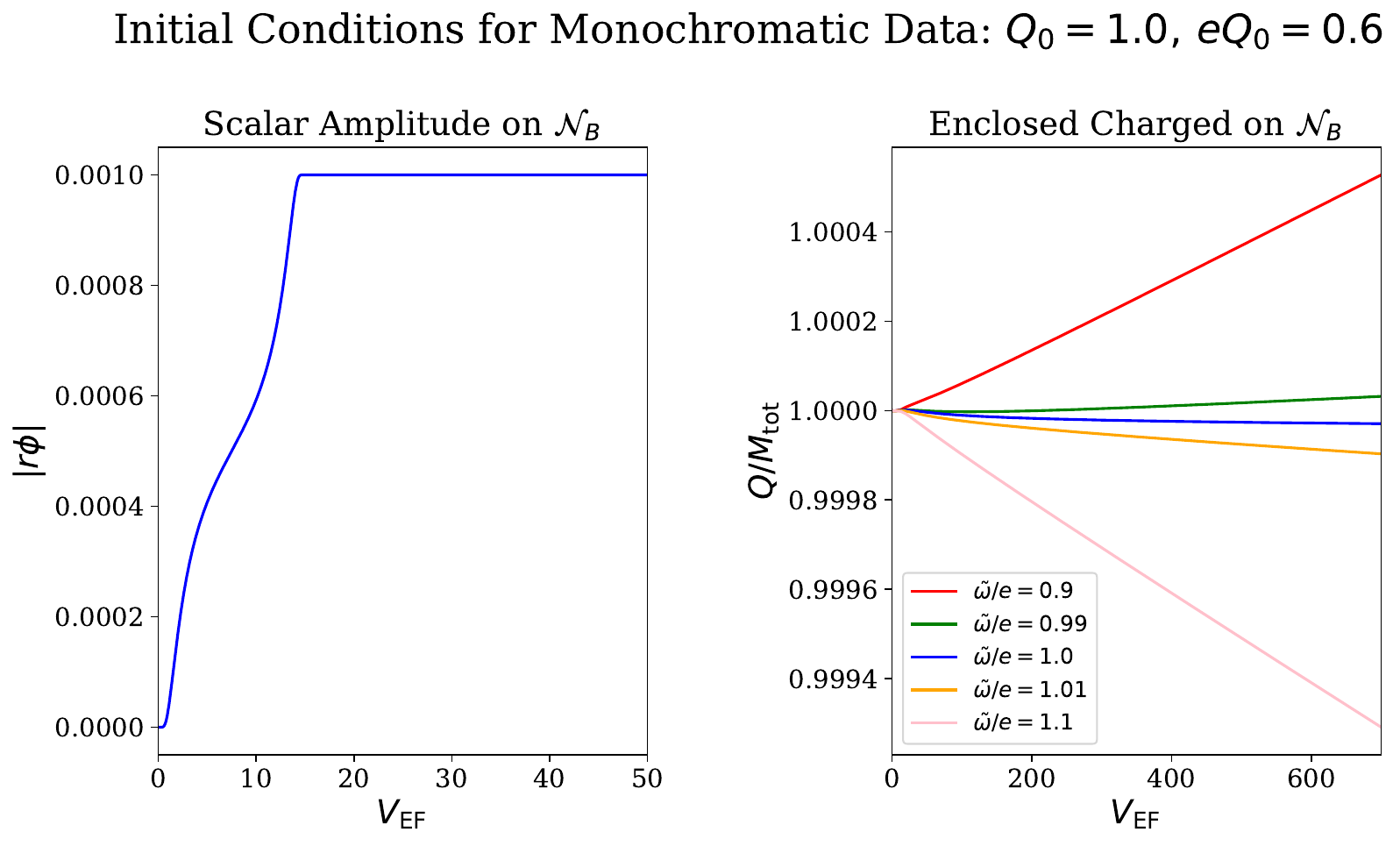}  
    \caption{Initial conditions for monochromatic data with $Q_0=1.0$, $\e Q_0=0.6$. The value of $\e/\tilde{\omega}$ is directly interpreted as the charge-to-mass ratio on $\mathcal{N}_B$.} 
    \label{fig:initsuperfig}
\end{figure*}

Initial data for the extremal background with strong charge coupling ($\e Q_0=0.6$) is shown in Figure~\ref{fig:initsuperfig}. Since the amplitude of the scalar field is constant at late times, we can \emph{directly} interpret the quantity $\e/\tilde{\omega}$ as a charge-to-mass ratio. This explains why all three curves on the right panel of Figure~\ref{fig:initsuperfig} have a (roughly) constant slope past $V_{\rm EF}=15$. In particular, the curve with $\e/\tilde{\omega}=1$ has constant $Q/M_{\rm tot}$ at late times. 

\begin{figure*}[t]
    \centering
    \includegraphics[width=0.99\textwidth]{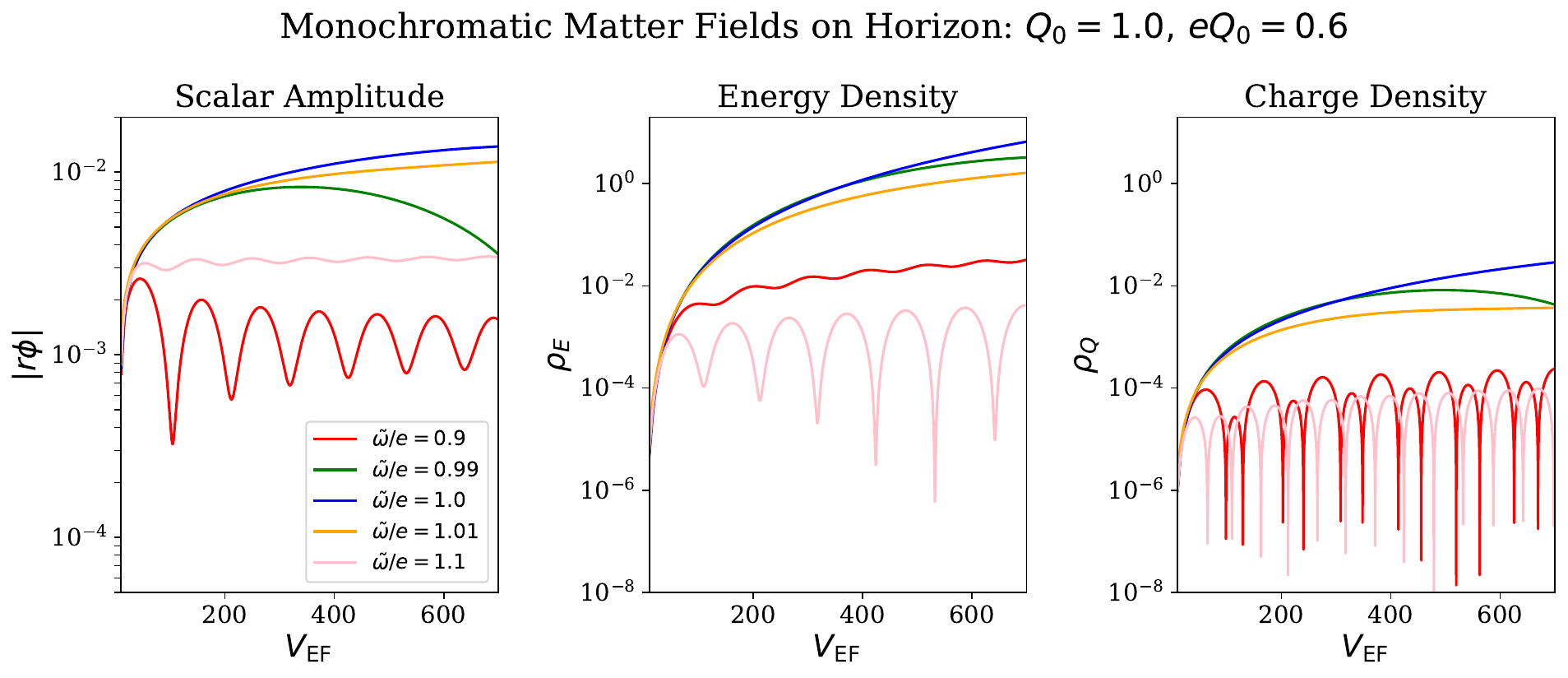}  
    \caption{Evolution of the matter fields on the future horizon of an extremal black hole. Initial data is monochromatic (modulo an early transient) with Lorenz gauge frequency $\tilde{\omega}$ and coupling $\e Q_0=0.6$.} 
    \label{fig:superfig}
\end{figure*}

After numerically evolving these fields to the horizon, we obtain the results shown in Figure~\ref{fig:superfig}. As one can see, the scalar amplitude $|r\phi|$ appears to turn over and begin oscillating for $\tilde{\omega}\neq \e$, with the turnover occurring more gradually as $\tilde{\omega}$ approaches $\e$. When $\tilde{\omega}=\e$ identically, a turnover is not apparent at all. Moreover, the energy and charge densities grow the fastest along the horizon for $\tilde{\omega}=\e$ as well. 

We note that at larger $V_{\rm EF}$ than what is depicted in Figure~\ref{fig:superfig}, the $\tilde{\omega}=\e$ curve will indeed begin to turn over and oscillate once the nonlinear effects of SED kick in. Here, we are limiting our study to small amplitudes, hence suppressing these non-linear terms until very late times so that the disconnect between $Q_0$ and $Q_f$ remains small. However, we have confirmed that as the initial data amplitude is increased, the $\tilde{\omega}=\e$ curve begins to oscillate sooner. 

At extremality, the limiting value of the NZDM frequency coincides with the onset of the {\em superradiance} phenomenon, where energy can be extracted from the black hole. Superradiant scattering of charged scalar waves off RN black holes is analogous to that of gravitational waves scattering off Kerr black holes, except the relevant property governing the process is charge as opposed to angular momentum \cite{brito_superradiance_2020}. Namely, modes with ${\rm Re}(\omega)>{\rm Re}(\omega_{\rm SR})$ inject charge into the black hole, whereas modes with ${\rm Re}(\omega)<{\rm Re}(\omega_{\rm SR})$ extract charge from the black hole. In Lorenz gauge
\cite{hod_2010_v2,hod_stability_2012,brito_superradiance_2020}\begin{align}
\label{eq:ZDMdef}
   \omega_{\rm SR}|_{\rm Lorenz}&=\frac{\e Q_0}{r_+}.
\end{align}
At extremality, we have $\omega_{\rm SR}|_{\rm Lorenz}=\omega_{\rm NZD}|_{\rm Lorenz}$.

This connection between the NZDM and the superradiant bound frequency provides an intuitive and physical explanation for the {\em enhancement} of the Aretakis instability in the presence of electromagnetic coupling $\e \neq 0$, as follows. At the exact onset of superradiance, the electromagnetic flux through the horizon drops to zero (at the linear level), effectively shutting off a channel of energy/charge loss for horizon perturbations. Moreover, the FFT shown in Fig. \ref{fig:fftfig} suggests this frequency is strongly excited in  perturbations more generic than the driven problem studied in this section. So in cases where the NZDM coincides with the onset of superradiance, which by itself further suppresses decay, it is not surprising that we see an enhanced Aretakis instability.

\begin{figure*}[t]
    \centering
    \includegraphics[width=0.99\textwidth]{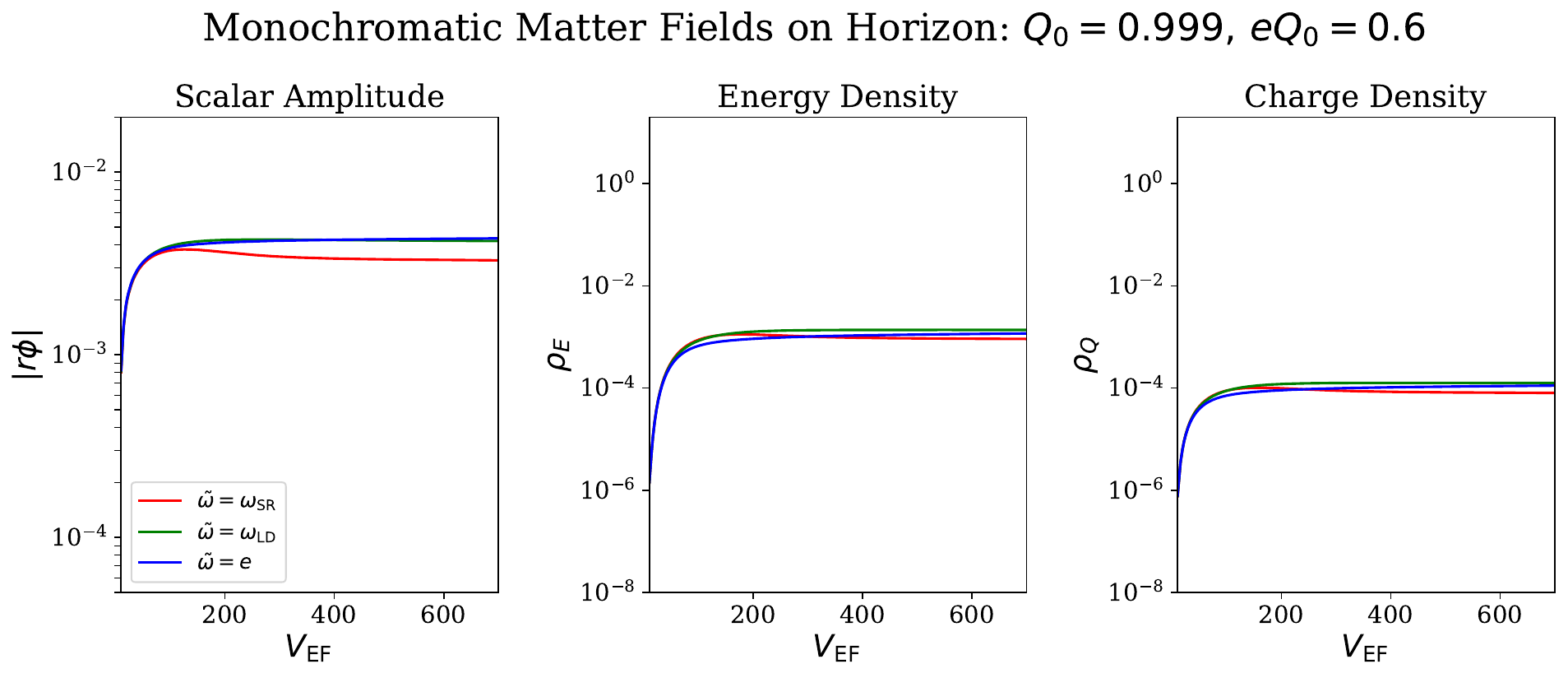}  
    \caption{Evolution of the matter fields on the future horizon of a sub-extremal black hole ($Q_0=0.999$). Initial data is monochromatic (modulo an early transient) with Lorenz-gauge frequency $\tilde{\omega}$ and coupling $\e Q_0=0.6$. The scale on all three panels matches that of Fig.~\ref{fig:superfig}. Here, $\omega_{\rm SR}=0.956\e$ is the superradiant bound frequency in Lorenz gauge, and $\omega_{\rm LD}=0.977\e$ is the real part of the least-damped QNM frequency in Lorenz gauge.} 
    \label{fig:subsuperfig}
\end{figure*}

For comparison, we also illustrate the monochromatic scattering problem for the sub-extremal case ($Q_0=0.999$) in Figure~\ref{fig:subsuperfig}. Away from extremality, the superradiant bound frequency ($\tilde{\omega}=0.956\e$), least damped QNM frequency $(\tilde{\omega}=0.977\e$), and frequency producing a charge-to-mass ratio of unity $(\tilde{\omega}=\e)$ are all distinct. However, one can see in Figure~\ref{fig:subsuperfig} that for all three frequencies, the physical observables asymptote to constants on the horizon. This behavior is expected when driving a damped system with a constant-amplitude wave and stands in clear contrast to the extremal case when driven at the NZDM (superradiant bound) frequency. 

Finally, we remark that our numerical simulations with monochromatic initial data still support the claim that $\tilde{\omega}=\e$ is a ``nearly-zero-damped mode'' rather than a genuine ``zero-damped mode.'' While the growth we see is akin to driving an undamped mode at its resonant frequency, we find in contrast that if the incoming, monochromatic initial data is (smoothly) turned off at some late time $V_{\rm off}$, the resulting scalar amplitude on the horizon breaks to power-law decay shortly after $V=V_{\rm off}$. If this were a genuinely ``zero-damped mode," its amplitude would have remained constant after turning off its driver.

\section{Summary and Discussion\label{sec:conclusion}}

In this paper, we have presented a numerical analysis of scalar electrodynamics on a fixed extremal RN background. By evolving the charged wave equation and Maxwell's equations in tandem, we have taken a step forward in understanding the fully non-linear endpoint of the Aretakis instability in this spherically symmetric setting.

In particular, we have developed a numerical code in compactified, double-null coordinates to evolve initial data to the event horizon and future null infinity. Our simulation results confirm earlier analyses of the linearized problem. They also fill in several power-law scaling relations for the behavior of the system on the horizon and at null infinity that, to our knowledge, have not been presented before in the literature (see Table~\ref{table:table1}).
Most notably, for strong charge coupling $(\e Q_0\geq 1/2)$, we find that, as measured by an observer free falling into the black hole from rest at infinity, the energy density on the horizon grows as
\begin{align}
    \rho_E\sim (\p_r|r\phi|)^2\sim V_{\rm EF}^1,
\end{align}
while the horizon charge density $\rho_Q$ asymptotes to a non-zero constant at late times in terms of the Eddington-Finkelstein null coordinate $V_{\rm EF}$. Our work also builds on the previous analytic studies by incorporating the full nonlinearities of Maxwell's equations. However, we consider only initial data with amplitude $\ll 1$ so that the nonlinearities of SED remain in check and that the disconnect between the dynamical charge $Q$ and the charge of the black hole remains small.

We have further given more insight into the physical nature of the Aretakis instability by exploring the response of the system driven by an incoming monochromatic wave. As argued in several works before \cite{zimmerman_horizon_2017,casals_horizon_2016,richartz_synch_2}, the Aretakis instability can be connected to the existence of nearly zero-damped modes (NDZMs) for sub-extremal black holes. The damping timescale of these modes goes to zero approaching extremality, yet at {\em exact} extremality they cease to exist as complex exponential modes, instead effectively weakening the late time power-law decay of the field on the horizon. Moreover, for the charged scalar field, the limiting frequency of the NZDMs coincides with the onset of charged superradiance, with further weakens the decay of the field. 

We have added to this picture by showing that when driving a sub-extremal RN black hole with a (strongly coupled) charged scalar wave with frequency at the onset of superradiance, the scalar amplitude on the horizon asymptotes to a constant. But at extremality, we observe unbounded growth of the scalar field until non-linearities from SED take over, indicative of driving a system at an undamped resonance. Again, we emphasize there are no normal (undamped) natural oscillation modes of the charged scalar field on extremal RN, though it seems plausible to connect the existence of a driven resonance with significantly weaker decay to generic perturbations, thus giving rise to the enhanced Aretakis instability.

In establishing these results, we have utilized a novel gauge invariant form of SED applied to this system. Although it is easier from a numerical perspective to evolve the gauge-dependent complex scalar $\phi$ and the vector potential $A_\mu$, it is more insightful to analyze only the gauge-invariant fields $P\equiv |r\phi|$ and $Q$. Regarding the gauge for numerical evolution, we have also introduced a {\em quasi-Lorenz} gauge (Eq.~\ref{eq:quasilorenz}), which has better asymptotic properties for studying RN perturbations than the more standard Lorenz condition.  

In the future, we plan to incorporate Einstein's equations into our numerical analysis. This will allow us to complete our extension of MRT \cite{murata_what_2013}, who studied the completely nonlinear evolution of an \emph{uncharged} scalar field in an asymptotically RN spacetime. They found that generically, metric backreaction pushes the black hole away from extremality at late times; however, for sufficiently fine-tuned initial data, one can construct a ``dynamical" extremal black hole. It will be interesting to explore how their conclusions change, if at all, when allowing for dynamics in both the charge and mass of the black hole. 

\section*{Acknowledgements}
We thank Mihalis Dafermos,  Suvendu Giri, Shahar Hadar, Christoph Kehle, Eliot Quataert, and Ryan Unger for helpful discussions. ZG is supported by a NSF Graduate Research Fellowship. FP acknowledges support from the NSF through the grant PHY-220728. Some of the simulations presented in this work were performed on computational resources managed and supported by Princeton Research Computing, a consortium of groups including the Princeton Institute for Computational Science and Engineering (PICSciE) and the Office of Information Technology's High Performance Computing Center and Visualization Laboratory at Princeton University.

\appendix
\begin{widetext}

\section{MRT Coordinates \label{app:coords}}
In this Appendix we expand on properties of the MRT coordinates, and we demonstrate their smoothness across both the sub-extremal and extremal outer horizons. As first expressed in Eq.~\ref{eq:murataeq}, the compactified MRT coordinates $\{u,v\}$ are defined implicitly in terms of the EF null coordinates $\{U_{\rm EF},V_{\rm EF}\}$ as\begin{align}
    U_{\rm EF}&=-2r_\star(r_+-\tan u/2),\quad V_{\rm EF}=2r_\star(r_++\tan v/2),
\end{align}
and again note that quantities in parenthesis are arguments of the function $r_\star(r)$.
The radial coordinate $r$ can thus be expressed as a function of $u$ and $v$ via\begin{align}
    r_\star(r)&=r_\star(r_+-\tan u/2)+r_\star(r_++\tan v/2).
\end{align}
To see that these relations remain well-defined at the outer horizon, we follow \cite{murata_what_2013} (cf. their Equation 63),
and Taylor expand the above relation near $r=r_+$, giving\begin{align}
    r(u,v)&=\begin{cases}
        r_+-\frac{1}{2}e^{\kappa_+V_{\rm EF}}u+\mathcal{O}(u^2),&\text{Future horizon}\\
        r_++\frac{1}{2}e^{\kappa_+U_{\rm EF}}v+\mathcal{O}(v^2),&\text{Past horizon}.
        \end{cases}
\end{align}
The above expressions are smooth everywhere on the horizon and are also continuous in the extremal limit $\kappa_+\to 0$. Plugging the above expansion into Eq.~\ref{eq:muratametric} gives the following limiting expressions for the $UV$ component of the metric in MRT coordinates:\begin{align}
\label{eq:metriclimit}
    \lim_{u\to 0} f_{\rm MRT}&=\frac{e^{\kappa_+V_{\rm EF}}\sec^2v}{2F(r_++\tan v/2)},\qquad
    \lim_{v\to 0}f_{\rm MRT}=\frac{e^{\kappa_+U_{\rm EF}}\sec^2u}{2F(r_+-\tan u/2)}.
\end{align}
The above expressions are manifestly non-singular across the outer horizon, and are also continuous in the extremal limit $\kappa_+\to 0$. The compactified MRT coordinates are thus a horizon-penetrating coordinate system.

\section{Quasi-Lorenz Gauge\label{app:qlorenz}}
To show that Quasi-Lorenz gauge introduces sharper decay than Lorenz gauge, let us derive the components of the gauge field $A_\mu$ in Quasi-Lorenz gauge for a point charge $Q_0$ at the origin. In any choice of double-null coordinates $\{U,V\}$, it is easy to see that the gauge condition\begin{align}
    A_{U,V}+A_{V,U}=0
\end{align}
combined with Eq.~\ref{eq:faradaydef} 
has the solution\begin{align}
\label{eq:doubleAint}
    A_U(U,V)&=Q_0\int_{V_0}^VdV'\frac{f(U,V')}{2r^2(U,V')},\qquad A_V(U,V)=-Q_0\int_{U_0}^UdU'\frac{f(U',V)}{2r^2(U',V)},
\end{align}
which satisfies the boundary condition that $A_U(U,V_0)=A_V(U_0,V)=0$. Now, let us explicitly evaluate these integrals in \emph{un}compactified MRT coordinates, which we will 
refer to as $U$ and $V$ throughout the rest of this Appendix. In this case, we can change the integration measure to $dr$ in each integral via\begin{align}
    dV=\frac{dV}{dV_{\rm EF}}\frac{\p V_{\rm EF}}{\p r_\star}\bigg|_{U}\frac{dr_\star}{dr}dr=\frac{2F(r_++V/2)}{F(r)}dr,\quad dU=\frac{dU}{dU_{\rm EF}}\frac{\p U_{\rm EF}}{\p r_\star}\bigg|_{V}\frac{dr_\star}{dr}dr=-\frac{2F(r_+-U/2)}{F(r)}dr,
\end{align}
where the factor of 2 comes from\begin{align}
    \frac{\p V_{\rm EF}}{\p r_\star}\bigg|_{U}&=\left(\frac{\p r_\star}{\p V_{\rm EF}}\bigg|_{U}\right)^{-1}=\left[\p_V\left(\frac{V_{\rm EF}-U_{\rm EF}}{2}\right)\right]^{-1}=2
    \\
    \frac{\p U_{\rm EF}}{\p r_\star}\bigg|_{V}&=\left(\frac{\p r_\star}{\p U_{\rm EF}}\bigg|_{V}\right)^{-1}=\left[\p_U\left(\frac{V_{\rm EF}-U_{\rm EF}}{2}\right)\right]^{-1}=-2.
\end{align}
Thus, the integrals become\begin{align}
    A_{U}&=-\frac{Q_0}{2F(r_+-U/2)}\int_{r(U,V_0)}^{r(U,V)}\frac{dr'}{r'^2}=-\frac{Q_0}{2F(r_+-U/2)}\left[\frac{1}{r(U,V)}-\frac{1}{r(U,V_0)}\right]
    \\
    A_{V}&=-\frac{Q_0}{2F(r_++V/2)}\int_{r(U_0,V)}^{r(U,V)}\frac{dr'}{r'^2}=-\frac{Q_0}{2F(r_++V/2)}\left[\frac{1}{r(U,V)}-\frac{1}{r(U_0,V)}\right].
\end{align}
This result is valid everywhere except the outer horizon: the above expression for $A_{U}$ becomes ill-defined when $U=0$, and the above expression for $A_{V}$ becomes ill-defined when $V=0$. In that case, we revert to the original integrals in terms of the null coordinates (Eq.~\ref{eq:doubleAint}) and apply Eq.~\ref{eq:metriclimit} for the expression for $f_{\rm MRT}$ along the outer horizon, giving
\begin{align}
    \lim_{U\to 0}A_{U}&=\frac{Q_0}{4r_+^2}\int_{V_{0}}^{V}dV'\frac{e^{2\kappa_+r_\star(r_++V'/2)}}{F(r_++V'/2)}= \begin{cases}
        \frac{Q_0}{8r_+^3\kappa_+}\left[Ve^{\kappa_+V}\left(\frac{r_+}{V/2+r_+-r_-}\right)^{\kappa_+/\kappa_-}-V\leftrightarrow V_0\right],&|Q_0|<1
        \\
        Q_0\left[\left(\frac{\log V}{r_+}-\frac{1}{V}\right)-V\leftrightarrow V_0\right],&|Q_0|=1
    \end{cases}
    \\
    \lim_{V\to 0}A_{V}&=-\frac{Q_0}{4r_+^2}\int_{U_{0}}^{U}dU'\frac{e^{-2\kappa_+r_\star(r_+-U'/2)}}{F(r_+-U'/2)}=
    \begin{cases}
       - \frac{Q_0}{8r_+^3\kappa_+}\left[Ue^{-\kappa_+U}\left(\frac{r_+}{U/2+r_+-r_-}\right)^{\kappa_+/\kappa_-}-U\leftrightarrow U_0\right],&|Q_0|<1
        \\
       - Q_0\left[\left(\frac{\log U}{r_+}-\frac{1}{U}\right)-U\leftrightarrow U_0\right],&|Q_0|=1,
    \end{cases}
\end{align}
which is regular. Thus, the point-charge vector potential in this gauge is
\begin{align}
    A_{U}&=\begin{cases}
        \frac{Q_0}{2F(r_+-U/2)}\left[\frac{1}{r(U,V_0)}-\frac{1}{r(U,V)}\right],&U\neq 0
        \\
        \frac{Q_0}{8r_+^3\kappa_+}\left[Ve^{\kappa_+V}\left(\frac{r_+}{V/2+r_+-r_-}\right)^{\kappa_+/\kappa_-}-V\leftrightarrow V_0\right],&U=0,|Q_0|<1
        \\
        Q_0\left[\left(\frac{\log V}{r_+}-\frac{1}{V}\right)-V\leftrightarrow V_0\right],&U=0,|Q_0|=1
    \end{cases}
    \\
    A_{V}&=\begin{cases}
        \frac{Q_0}{2F(r_++V/2)}\left[\frac{1}{r(U_0,V)}-\frac{1}{r(U,V)}\right],&V\neq 0
        \\
        -\frac{Q_0}{8r_+^3\kappa_+}\left[Ue^{-\kappa_+U}\left(\frac{r_+}{U/2+r_+-r_-}\right)^{\kappa_+/\kappa_-}-U\leftrightarrow U_0\right],&V=0,|Q_0|<1
        \\
        -Q_0\left[\left(\frac{\log U}{r_+}-\frac{1}{U}\right)-U\leftrightarrow U_0\right],&V=0,|Q_0|=1.
    \end{cases}
\end{align}
While these expressions may look like they decay as $r^{-1}$, the terms $r(U,V)^{-1}$ and $r(U,V_0)^{-1}$ cancel to leading order at null infinity, thus leaving a vector potential that decays as $r^{-2}$. 

As such, the compactified vector potential will remain finite at null infinity. In terms of compactified MRT coordinates $\{u,v\}$, we have\begin{align}
    A_u=\sec^2u A_{U},\qquad A_v=\sec^2v A_{V}.
\end{align}
Taylor expanding the uncompactified vector potential in large $r$ to find the first-order result, we find:\begin{align}
    A_u|_{\mathcal{I}^-}&=\lim_{U\to-\infty}U^2A_{U}=Q_0(V_{\rm EF}-V_{\rm EF,0})
    \\
    A_v|_{\mathcal{I}^+}&=\lim_{V\to\infty}V^2A_{V}=Q_0(U_{\rm EF}-U_{\rm EF,0}).
\end{align}
Thus, the compactified MRT coordinates produce a simple, finite, and smooth prescription for the vector potential at null infinity.

We remark that a different choice of gauge (based on conformal slicing) that is well-behaved at null infinity is also used in \cite{baake_superradiance_2016}; see their Appendix B for additional discussion of gauges.

\section{Gauge-Invariant Equations of Motion}
\label{app:gaugeinveom}
Here, we re-cast the equations of motion for SED 
in a manifestly gauge-invariant form. This will yield Eqs.~\ref{eq:peq}-\ref{eq:qeq}. As in \S\ref{sec:eom}, we begin by defining a scalar amplitude $P\equiv |r\phi|$, which is related to the fields $\overline{\xi}$ and $\overline{\Pi}$ via\begin{align}
    (\overline{\xi},\overline{\Pi})=(P\cos\alpha,P\sin\alpha),
\end{align}
with $\alpha$ the local phase of the complex field $r\phi$. If we differentiate the expression for $P$ and apply the equations of motion for $\overline{\xi}$ and $\overline{\Pi}$, we obtain:\begin{align}
     \label{eq:puv1}P_{,UV}&=P\left[g+\e^2\overline{A}_U\overline{A}_V\right],
\end{align}
where we have introduced the gauge-invariant potential\begin{align}
    \overline{A}_\mu\equiv A_\mu-\frac{\alpha_{,\mu}}{\e}.
\end{align}
From here, we write down Maxwell's equations in terms of these gauge-invariant quantities. The current $J_\mu$ simplifies dramatically:\begin{align}
    J_\mu&=-\frac{2\e^2P^2\overline{A}_\mu}{r^2},
\end{align}
which turns Maxwell's constraint equations (Eqs.~\ref{eq:maxwellcons1}-\ref{eq:maxwellcons2}) into\begin{align}
    Q_{,U}&=-8\pi \e^2P^2\overline{A}_{U},\qquad Q_{,V}=8\pi \e^2P^2\overline{A}_V.
\end{align}
These can be substituted into Eq.~\ref{eq:puv1} to yield\begin{align}
    P_{,UV}=Pg-\frac{Q_{,U}Q_{,V}}{64\pi^2\e^2P^3},
\end{align}
which matches Eq.~\ref{eq:peq}.

Next, we can derive a gauge-invariant evolution equation for $Q$ by differentiating Maxwell's constraint equations:\begin{align}
    Q_{,UV}&=\frac{1}{2}[(4\pi r^2J_U)_{,V}-(4\pi r^2J_V)_{,U}]=8\pi \e^2P\left[\overline{A}_VP_{,U}-\overline{A}_{U}P_{,V}-\frac{fPQ}{2r^2}\right]
    \\&=
    \frac{P_{,U}Q_{,V}}{P}+\frac{P_{,V}Q_{,U}}{P}-\frac{4\pi \e^2fP^2Q}{r^2},
\end{align}
which matches Eq.~\ref{eq:qeq}.

While the equations derived above are written in terms of $U$ and $V$, they can be put into a more general covariant form:\begin{align}
    r\nabla_\mu \nabla^\mu (Pr^{-1})=-\frac{\nabla_\mu Q\nabla^\mu Q}{64\e^2\pi^2 P^3},\qquad r^2\nabla_\mu\nabla^\mu Q&=8\pi \e^2P^2Q+\frac{2r\nabla_\mu Q\nabla^\mu (rP)}{P},
\end{align}
which, as one can readily check, reduces to Eqs.~\ref{eq:peq}-\ref{eq:qeq} upon choice of a double-null coordinate system.

\section{Numerical Evolution Scheme\label{app:numerics}}
In this Appendix, we detail our numerical evolution scheme and give convergence results.

\subsection{PDE Discretization}
In our numerical scheme, we first discretize the spacetime into a grid of cells in $U$ and $V$, as depicted in the Penrose diagram in Figure~\ref{fig:numericpenrose}.
\begin{figure}[h]
    \centering
    \includegraphics[width=0.5\textwidth]{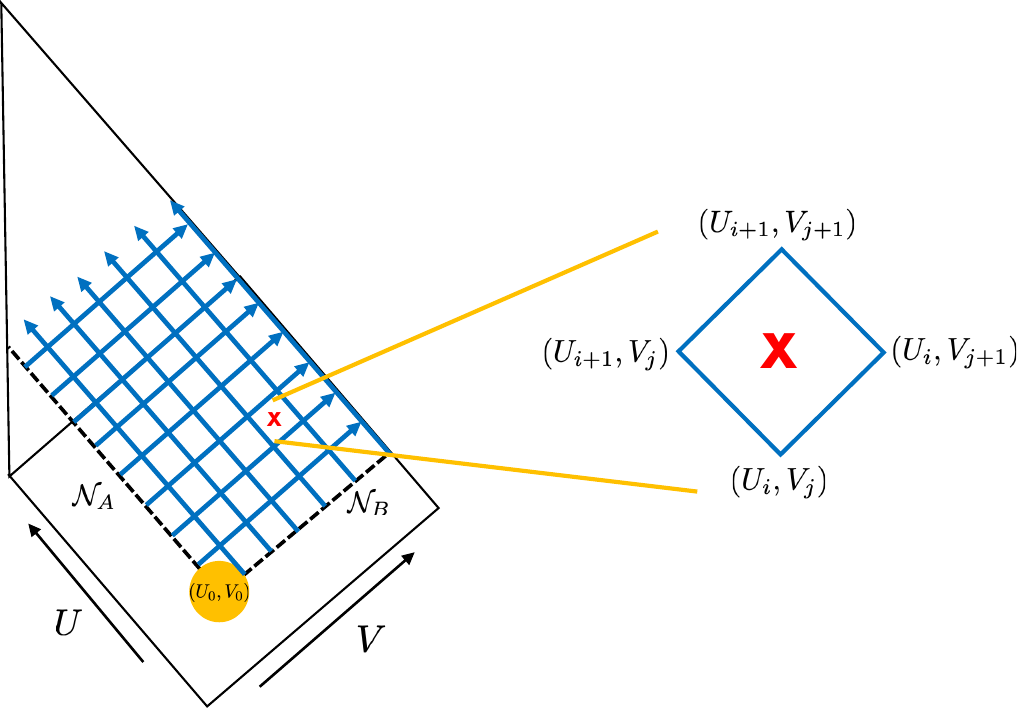}  
    \caption{Penrose diagram for Extremal RN spacetime, superposed with a discretized mesh in $U$ and $V$. The initial data origin $(U_0,V_0)$ is demarcated with an orange disk, and the hypersurfaces $\mathcal{N}_A$ and $\mathcal{N}_B$ are demarcated as dashed lines. One grid cell is amplified, which has its $UV$ coordinates labeled with an ``X" to mark its center.} 
    \label{fig:numericpenrose}
\end{figure}

To discretize the equations of motion (Eqs~\ref{eq:waveeqn}-\ref{eq:Aeq}), we employ the method of Burko \& Ori \cite{burko_late-time_1997}, which was also implemented in \cite{murata_what_2013,redondo2024ringdown}. This method exploits the fact that the derivatives of the fields at the \emph{center} of each cell can be expressed --- to second order --- in terms of the values of the fields at the \emph{corners} of each cell. Defining the corners of each cell as in Figure~\ref{fig:numericpenrose}, we can express all relevant derivatives of a field $\Phi$ as\begin{align}
\label{eq:disc1}
\Phi_{,U}|_{\rm X}&=\frac{(\Phi_{i+1,j}-\Phi_{i,j+1})+(\Phi_{i+1,j}-\Phi_{i,j+1})}{2\Delta U}+\mathcal{O}(\Delta U\Delta V)
\\
\Phi_{,V}|_{\rm X}&=\frac{(\Phi_{i,j+1}-\Phi_{i,j})+(\Phi_{i+1,j+1}-\Phi_{i+1,j})}{2\Delta V}+\mathcal{O}(\Delta U\Delta V)
\\
\Phi_{,UV}|_{\rm X}&=\frac{(\Phi_{i+1,j+1}-\Phi_{i,j+1})-(\Phi_{i+1,j}-\Phi_{i,j})}{\Delta U\Delta V}+\mathcal{O}(\Delta U\Delta V)
\\
\label{eq:disc2}
    \Phi|_{\rm X}&=\frac{\Phi_{i,j}+\Phi_{i+1,j}+\Phi_{i,j+1}+\Phi_{i+1,j+1}}{4}+\mathcal{O}(\Delta U\Delta V),
\end{align}
where ``X" means at the center of the cell, and $\Delta U\times \Delta V$ are the dimensions of the cell.

Starting at the origin $(U_0,V_0)$, we integrate from $\mathcal{N}_A$ to future null infinity along each hypersurface of constant $U$. At each cell we encounter, the values of the dynamical fields will be known at all corners of the cell except the top corner $(i+1,j+1)$. By employing Eqs.~\ref{eq:disc1}-\ref{eq:disc2} to evaluate the discretized system of PDE's at the center of the cell, we then apply the Newton-Raphson method to solve for the fields at the top corner.

The equations of motion that we use in the numerical scheme match those of Eq.~\ref{eq:waveeqn}-\ref{eq:Aeq}, with one crucial modification. We find that Maxwell's constraints are better preserved if we re-write gradients of the vector potential directly in terms of the charge $Q$:\begin{align}
    A_{U,V}&=\frac{Qf}{2r^2},\qquad A_{V,U}=-\frac{Qf}{2r^2},
\end{align}
with $Q$ then getting its own evolution equation:\begin{align}
    Q_{,UV}&=2\pi r^2(J_{U,V}-J_{V,U})=8\pi \e\bigg[-\frac{\e fQ(\overline{\xi}^2+\overline{\Pi}^2)}{2r^2}-\e A_U(\overline{\xi}\overline{\xi}_{,V}+\overline{\Pi}\overline{\Pi}_{,V})+\e A_V(\overline{\xi}\overline{\xi}_{,U}+\overline{\Pi}\overline{\Pi}_{,U})-\overline{\xi}_{,U}\overline{\Pi}_{,V}+\overline{\xi}_{,V}\overline{\Pi}_{,U}
    \bigg].
\end{align}
Thus, there are five evolution equations we simultaneously solve at each cell, with one for each dynamical field: $\{\overline{\xi},\overline{\Pi},A_U,A_V,Q\}$. These evolution equations are solved explicitly in terms of compactified MRT coordinates $\{u,v\}$, which reach both null infinity and the event horizon in finite coordinate time. 

However, the metric components $r$ and $f_{\rm MRT}$ both blow up at null infinity, leading to potential numerical issues. We discuss how to remedy these issues below. 

\subsection{Coordinate Renormalization}
To ensure that the equations of motion remain well-defined near null infinity, we find it useful to factor out the divergences and introduce a ``renormalized" $uv$-metric component and radial coordinate:\begin{align}
    \tilde{f}\equiv f\cos^2u\cos^2v,\quad R\equiv \frac{2r}{\tan v-\tan u}.
\end{align} 
These remain positive and well-defined throughout the spacetime. Minkowski space, for example is defined by\begin{align}
    \tilde{f}_{\rm Mink}=\frac{1}{2},\qquad R_{\rm Mink}=1,
\end{align}
and the RN solution also has $R_{\rm RN}=1$ at null infinity. The only point in the simulation domain where $\tilde{f}$ is undefined is at timelike infinity: $(u,v)=(0,\pi/2)$. This is the point labeled $i^+$ in the Penrose diagram of Figure~\ref{fig:penrosediagram}. As explained in \S\ref{sec:RNsec}, this point is where the future horizon, Cauchy horizon, and null infinity all ``meet''. In integrating the equations of motion, we can therefore only approach this point asymptotically.

It is straightforward to recast the evolution equations in terms of $\tilde{f}$ and $R$. While factors like $\sec^2v$ still appear in the equations, these will never need to be evaluated at $v=\pi/2$ exactly; null infinity lives along the edge of the outermost grid cells, but we evaluate the discretized PDE's only at the centers of the cells, where $u$ and $v$ remain finite.

\subsection{Non-uniform mesh spacing}
We have emperically found it advantageous to use the following non-uniform mesh structure for evolving
the sub-extremal $Q_0=0.999$ and extremal $Q_0=1$ cases studied in this paper.

For the sub-extremal grid, we space cells evenly in $V_{\rm EF}$ from $V_{\rm EF}=V_{\rm EF,0}$ to $V_{\rm EF}=1000$, and then logarithmically from $V_{\rm EF}=1000$ to $V_{\rm EF}=10000$, before adding a single point at $v_{\rm MRT}=\pi/2$ (null infinity). In the exterior, we then space evenly in $U_{\rm EF}$ from $U_{\rm EF}=U_{\rm EF,0}$ to $U_{\rm EF}=950$, before adding a single point at $u_{\rm MRT}=0$ (the horizon) and then reflecting the grid across the horizon into the black hole interior. Due to the exponential sensitivity of the horizon redshift effect (Eq.~\ref{eq:expo}), we cannot go past $U_{\rm EF}=950$ before hitting machine precision (we use quadruple precision floating point arithmetic).

For the extremal grid, we use the same spacing in $V$. But for $U$, we space evenly in $U_{\rm EF}$ from $U_{\rm EF}=U_{\rm EF,0}$ to $U_{\rm EF}=1500$, and then logarithmically from $U_{\rm EF}=1500$ to $U_{\rm EF}=10^5$, before adding a single point at $u_{\rm MRT}=0$ (the horizon) and reflecting the grid across the horizon into the interior. Since there is no redshift effect of the extremal horizon, we can to push much larger values of $U_{\rm EF}$ with quadruple precision floating point arithmetic and still obtain convergent solutions.

\subsection{Convergence}
Here, we demonstrate convergence of our code. To do so, we need to compare the code output at (at least) three different grid resolutions. Since we use a non-uniform mesh, the way we define ``different'' resolutions here is we choose one specific mesh as our base ``low'' resolution mesh, then define each successively higher resolution meshes as the immediately lower resolution mesh with each cell cut into quarters (as measured in compactified MRT coordinates). In other words, we double the resolution at each step. Consequently, for a second order accurate evolution scheme as we employ, in the convergent regime we expect global truncation error to drop by a factor of four per doubling step.

For our low resolution mesh, we construct it as described above with the exterior portion of the mesh having $20000$ cells in the $V$ direction and $10000$ cells in the $U$ direction. We then runs simulations at twice and four times this resolution. For some quantity $\Phi$ output from these simulations, we compute the point-wise convergence parameter $\mathcal{C}$ (see e.g. \cite{choptuik_lectures_2006_v2})\begin{align}
\label{eq:cdef}
    \mathcal{C}\equiv \frac{\Phi_0-\Phi_1}{\Phi_1-\Phi_2},
\end{align}
where $\Phi_0$ is the field from the low resolution run, with $\Phi_{1}$,$\Phi_2$ getting progressively finer. For second order convergence, we expect $\mathcal{C}\to 4$.

 The convergence factor $\mathcal{C}$ on the future horizon and at null infinity is plotted in Figure~\ref{fig:conv1} for a representative set of parameters.

\begin{figure}[h]
    \centering
    \includegraphics[width=0.99\textwidth]{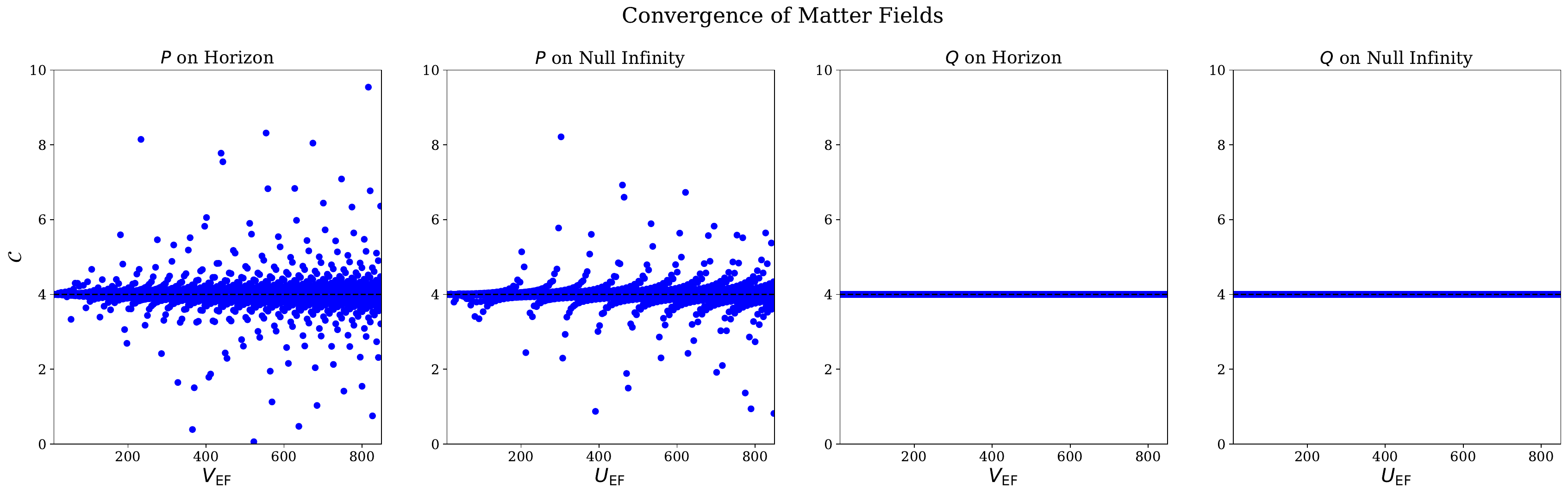}  
    \caption{Convergence factor for $P$ and $Q$ at the boundaries of the black hole exterior for the numerical simulation with $Q_0=1.0$, $\e Q_0=0.6$, and $\tilde{\omega}=0.0$. The idealized value $\mathcal{C}= 4$ is shown as a black dashed line in each panel.} 
    \label{fig:conv1}
\end{figure}

We see that in all cases, $\mathcal{C}$ is scattered around its idealized value of $4$, with the scatter in that of $Q$ typically being significantly smaller than that of $P$ at a fixed resolution. 

Next, we also demonstrate that Maxwell's constraints are preserved in our simulations. In Figure~\ref{fig:conv2}, we plot Maxwell's constraint equations:\begin{align}
    Q_{,u}-4\pi r^2J_u,\qquad Q_{,v}+4\pi r^2J_v,
\end{align}
which should each converge to zero quadratically with increasing resolution.
\begin{figure}[h]
    \centering
    \includegraphics[width=0.99\textwidth]{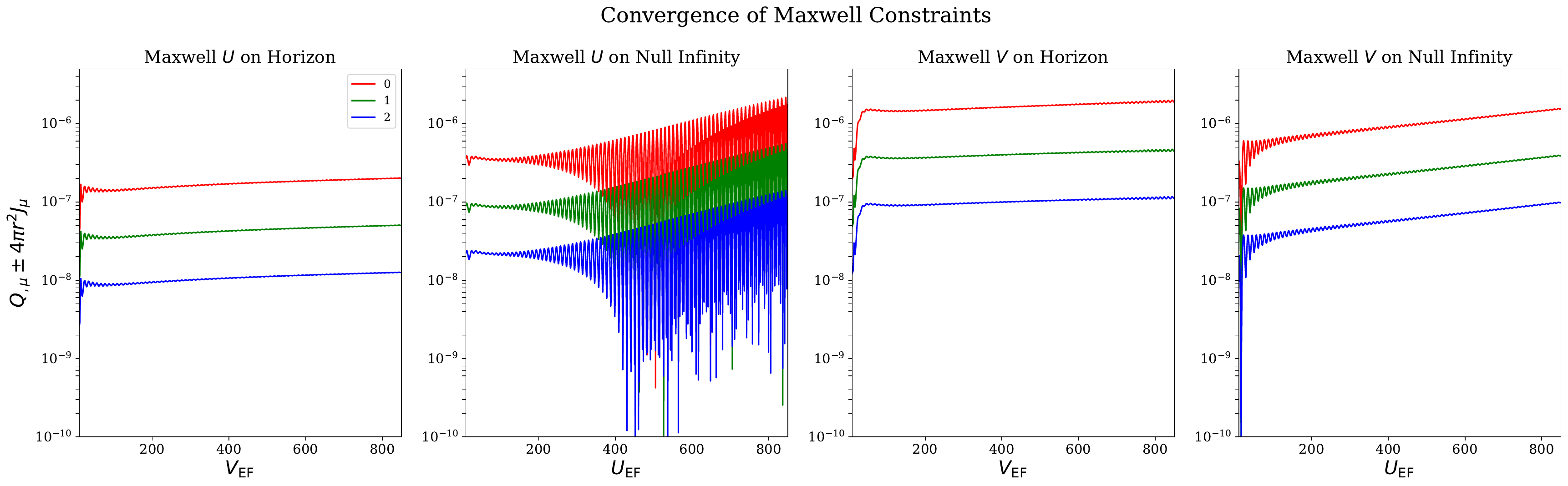}  
    \caption{Convergence of Maxwell's constraint equations on the boundaries of the black hole exterior, shown here for the simulation with $Q_0=1.0$, $\e Q_0=0.6$, and $\tilde{\omega}=0.0$. 
     ``Maxwell $U$" refers to the quantity $Q_{,U}-4\pi r^2J_U$, while ``Maxwell $V$" refers to the quantity $Q_{,V}+4\pi r^2J_V$. The curves (0,1,2) correspond to progressively increasing simulation resolution. }
    \label{fig:conv2}
\end{figure}
From the figure, we see that the trends with resolution show that both constraint equations are indeed satisfied to within truncation error on the plotted null hypersurfaces. 

\end{widetext}
\bibliographystyle{apsrev4-2} 
\bibliography{bib,references}{}
\end{document}